\documentclass[conference]{IEEEtran}

\IEEEoverridecommandlockouts

\usepackage{amsmath,amssymb,amsfonts}
\newtheorem{prop}{Proposition}
\usepackage{graphicx}
\usepackage{textcomp}
\usepackage{xcolor}
\usepackage{wasysym}
\usepackage{marvosym}
\usepackage{pifont}
\usepackage{centernot}
\usepackage{esvect}
\usepackage{hyperref}
\usepackage{enumerate}
\usepackage{algorithm}
\usepackage{proof}
\usepackage{mathtools}
\usepackage[noend]{algpseudocode}

\usepackage{listings}
\usepackage{esvect}
\usepackage{arydshln}

\usepackage{multirow}
\makeatletter
\newcommand{\linebreakand}{%
  \end{@IEEEauthorhalign}
  \hfill\mbox{}\par
  \mbox{}\hfill\begin{@IEEEauthorhalign}
}
\makeatother

\usepackage{listings, xcolor}

\definecolor{verylightgray}{rgb}{.97,.97,.97}

\lstdefinelanguage{Solidity}{
	keywords=[1]{anonymous, assembly, assert, 
	break, call, callcode, case, catch, class, constant, continue, constructor, contract, debugger, default, delegatecall, delete, do, else, emit, event, experimental, export, external, false, finally, for, function, gas, if, implements, import, in, indexed, instanceof, interface, internal, is,
	library, log0, log1, log2, log3, log4, memory, modifier, new, payable, pragma, private, protected, public, pure, push, require, return, returns, revert, selfdestruct, send, solidity, storage, struct, suicide, super, switch, then, this, throw, 
true, try, typeof, using, 
	view, while, with, addmod, ecrecover, keccak256, mulmod, ripemd160, sha256, sha3}, 
	keywordstyle=[1]\color{blue}\bfseries,
	keywords=[2]{address, bool, byte, bytes, bytes1, bytes2, bytes3, bytes4, bytes5, bytes6, bytes7, bytes8, bytes9, bytes10, bytes11, bytes12, bytes13, bytes14, bytes15, bytes16, bytes17, bytes18, bytes19, bytes20, bytes21, bytes22, bytes23, bytes24, bytes25, bytes26, bytes27, bytes28, bytes29, bytes30, bytes31, bytes32, enum, int, int8, int16, int24, int32, int40, int48, int56, int64, int72, int80, int88, int96, int104, int112, int120, int128, int136, int144, int152, int160, int168, int176, int184, int192, int200, int208, int216, int224, int232, int240, int248, int256, mapping, string, uint, uint8, uint16, uint24, uint32, uint40, uint48, uint56, uint64, uint72, uint80, uint88, uint96, uint104, uint112, uint120, uint128, uint136, uint144, uint152, uint160, uint168, uint176, uint184, uint192, uint200, uint208, uint216, uint224, uint232, uint240, uint248, uint256, var, void, ether, finney, szabo, wei, days, hours, minutes, seconds, weeks, years},	
	keywordstyle=[2]\color{teal}\bfseries,
	keywords=[3]{block, blockhash, coinbase, difficulty, gaslimit, number, timestamp, msg, data, gas, 
	sig, 
	now, tx, gasprice, origin},	
	keywordstyle=[3]\color{violet}\bfseries,
	identifierstyle=\color{black},
	sensitive=false,
	comment=[l]{//},
	morecomment=[s]{/*}{*/},
	commentstyle=\color{red}\ttfamily,
	stringstyle=\color{red}\ttfamily,
	morestring=[b]',
	morestring=[b]"
}

 \newcommand{\hide}[1]{}

\newcommand{\zap}[1]{}

\newcommand{\myset}[1]{\{ #1 \}}

\newtheorem{example}{Example}

\newcommand{\todoc}[2]{{\textcolor{#1} {\textbf{[[#2]]}}}}

\newcommand{\todored}[1]{\todoc{red}{#1}}

\newcommand{\TODO}[1]{\todored{#1}}

\newcommand{\assumecmd}{{\it assume}}
\newcommand{\assertcmd}{{\it assert}}

\newcommand{\false}{{\it false}}
\newcommand{\true}{{\it true}}

\newcommand{\ifcmd}{{\it if}}
\newcommand{\whilecmd}{{\it while}}

\newcommand{\myskip}[1]{}
\newcommand{\inducmap}{\mu}
\newcommand{\post}{\mathsf{sp}}

\newcommand{\mytrans}{\rightsquigarrow}

\newcommand{\cmark}{\textcolor{blue}{\ding{51}}}
\newcommand{\xmark}{\textcolor{red}{\ding{55}}}%

\newcommand{\free}{\mathsf{FV}}

\newcommand{\genvc}{\textsc{GenVC}}

\newcommand{\verismart}{{\textsc{VeriSmart}}}
\newcommand{\mythril}{{\textsc{Mythril}}}
\newcommand{\manticore}{{\textsc{MantiCore}}}
\newcommand{\osiris}{{\textsc{Osiris}}}
\newcommand{\oyente}{{\textsc{Oyente}}}
\newcommand{\zeus}{{\textsc{Zeus}}}
\newcommand{\smtchecker}{{\textsc{SMTChecker}}}
\newcommand{\securify}{{\textsc{Securify}}}

\newcommand{\maian}{{\textsc{Maian}}}
\newcommand{\gasper}{{\textsc{Gasper}}}
\newcommand{\reguard}{{\textsc{ReGuard}}}

\newcommand{\mylabel}{\textit{Label}}
\newcommand{\fol}{\textsf{FOL}}

\newcommand{\Verify}{\textsc{Validator}}
\newcommand{\UpdateInv}{\textsc{Generator}}

\newcommand{\myparagraph}[1]{\setlength{\parindent}{7pt}\paragraph*{\textbf{{#1}}}}

\newcommand{\myappendix}[1]{\subsection{\textbf{#1}}}

\newcommand{\Alarms}{\textsf{\#Alarm}}
\newcommand{\FP}{\textsf{\#FP}}
\newcommand{\Q}{\textsf{\#Q}}
\newcommand{\FN}{\textsf{\#FN}}

\newcommand{\CVE}{\textsf{CVE}}
\newcommand{\Verified}{\textsf{Verified}}

\newcommand{\ti}{\psi}

\newcommand{\lentry}{{\it entry}}
\newcommand{\lexit}{{\it exit}}

\newcommand{\RefineTran}{\textsc{Tran}}
\newcommand{\RefineLoop}{\textsc{Loop}}
\newcommand{\nextloopinv}{\textsc{RefineL}}
\newcommand{\nexttraninv}{\textsc{RefineT}}

\newcommand{\mysigma}{{\sf sum}}
\newcommand{\myf}{{\it f}}

\begin{document}

\title{\verismart: A Highly Precise Safety Verifier for
Ethereum Smart Contracts
\thanks{To appear in the IEEE Symposium on Security \& Privacy, May 2020}
}

\author{\IEEEauthorblockN{Sunbeom So, Myungho Lee, Jisu Park, Heejo Lee, Hakjoo Oh$^*$
\thanks{$^*$Corresponding author: Hakjoo Oh, {hakjoo\_oh@korea.ac.kr}}}
\IEEEauthorblockA{
Department of Computer Science and Engineering\\
Korea University 
}
\footnote{$^*$corresponding author}
}

\myskip{
\author{
\IEEEauthorblockN{Sunbeom So}
\IEEEauthorblockA{
Korea University \\
sunbeom\_so@korea.ac.kr}
\and
\IEEEauthorblockN{Myungho Lee}
\IEEEauthorblockA{
Korea University\\
myungho\_lee@korea.ac.kr}
\and
\IEEEauthorblockN{Jisu Park}
\IEEEauthorblockA{
Korea University\\
jisu\_park@korea.ac.kr}
\and
\IEEEauthorblockN{Heejo Lee}
\IEEEauthorblockA{
Korea University\\
heejo@korea.ac.kr}
\and
\IEEEauthorblockN{Hakjoo Oh$^*$}
\IEEEauthorblockA{
Korea University\\
hakjoo\_oh@korea.ac.kr}
}
}
\maketitle


\begin{abstract}

  We present \verismart, a highly precise verifier for ensuring
  arithmetic safety of Ethereum smart contracts.  Writing safe smart
  contracts without unintended behavior is critically important
  because smart contracts are immutable and even a single flaw can
  cause huge financial damage. In particular, ensuring that
  arithmetic operations are safe is one of the most important and common
  security concerns of Ethereum smart contracts nowadays. 
  In response, several
  safety analyzers have been proposed over the past few years, but
  state-of-the-art is still unsatisfactory; no existing tools achieve
  high precision and recall at the same time, inherently limited to
   producing annoying false alarms or missing critical bugs.
  By contrast, \verismart~aims for an uncompromising analyzer that
  performs exhaustive verification without compromising precision or
  scalability, thereby greatly reducing the burden of manually
  checking undiscovered or incorrectly-reported issues.  To achieve
  this goal, we present a new domain-specific algorithm for verifying
  smart contracts, which is able to automatically discover and
  leverage transaction invariants that are essential for precisely
  analyzing smart contracts.  Evaluation with real-world smart
  contracts shows that \verismart~can detect all arithmetic bugs with a
  negligible number of false alarms, far outperforming existing
  analyzers. 

\end{abstract}





\section{Introduction}

Safe smart contracts are indispensable for trustworthy blockchain
ecosystems.  Blockchain is widely recognized as one of the most
disruptive technologies and smart contracts lie at the heart of this
revolution (e.g.,~\cite{insurance,
  DBLP:journals/corr/abs-1806-00555}).  Smart contracts are computer
programs that run on blockchains in order to automatically fulfill
agreed obligations between untrusted parties without
intermediaries. Unfortunately, despite their potential, smart
contracts are more likely to be vulnerable than traditional programs
because of their unique characteristics such as openness and
immutability~\cite{Atzei:2017:SAE:3080353.3080363}.  As a result,
unsafe smart contracts are prevalent and are increasingly becoming a
serious threat to the success of the blockchain technology.  For
example, recent infamous attacks on the Ethereum blockchain such as
the DAO~\cite{dao} and the Parity Wallet~\cite{parity} attacks were
caused by unsafe smart contracts.

In this paper, we present \verismart, a fully automated safety
analyzer for verifying Ethereum smart contracts with a particular
focus on arithmetic safety. We focus on detecting arithmetic bugs such
as integer over/underflows and division-by-zeros because smart
contracts typically involve lots of arithmetic operations and they are
major sources of security vulnerabilities nowadays. For example,
arithmetic over/underflows account for 95.7\% (487/509) of CVEs
assigned to Ethereum smart contracts, as shown in
Table~\ref{table:cve}.  Even worse, arithmetic bugs, once exploited,
are likely to cause significant but unexpected financial damage (e.g.,
the integer overflow in the SmartMesh contract~\cite{peckshield-smt}
explained in
Section~\ref{sec:overview}).  
Our goal is to detect all arithmetic bugs 
before deploying smart contracts on the blockchain.


\begin{table}[t]
  \caption{Statistics on CVE-reported security vulnerabilities of
    Ethereum smart contracts (as of May. 31, 2019)}
  \setlength{\tabcolsep}{.2em}
\centering

\begin{tabular}{|c|c|c|c|c|c|}
    \hline
    Arithmetic 		& Bad 	  & Access & Unsafe Input &
                                                 								\multirow{2}{*}{Others} & \multirow{2}{*}{Total} \\[-0.1em]

 Over/underflow &  Randomness & Control & Dependency  		& &    \\ \hline
    487 (95.7 \%)       & 10 (1.9 \%)    & 4 (0.8 \%)        & 4 (0.8 \%)      & 4 (0.8\%)   & \multicolumn{1}{c|}{509}       \\ \hline
  \end{tabular}
  \label{table:cve}
   \vspace{-0.5em}
\end{table}

Unlike existing techniques,
\verismart~aims to be a truly practical tool by performing automatic,
scalable, exhaustive, yet highly precise verification of smart contracts. 
Recent years have seen an increased
interest in automated tools for analyzing arithmetic safety of smart
contracts~\cite{DBLP:conf/acsac/TorresSS18,mythril,oyente,manticore,
  DBLP:conf/ndss/KalraGDS18,DBLP:conf/isola/AltR18}.  However,
existing tools are still unsatisfactory. 
A major weakness of bug-finding approaches (e.g.,
\cite{DBLP:conf/acsac/TorresSS18,oyente,mythril,manticore}) is that
they are likely to miss fatal bugs (i.e., resulting in false
negatives), because they do not consider all the possible behaviors of the program.
On the other hand, verification approaches (e.g.,
\cite{DBLP:conf/ndss/KalraGDS18,DBLP:conf/isola/AltR18}) are exhaustive and therefore miss
no vulnerabilities, but they
typically do so at the expense of precision (i.e., resulting in false
positives). In practice, both false negatives and positives burden
developers with error-prone and time-consuming process for manually
verifying a number of undiscovered issues or incorrectly reported
alarms. 
\verismart~aims to overcome these shortcomings of existing
approaches by being exhaustive yet precise.  

To achieve this goal, we present a new verification algorithm 
for smart contracts.  The key feature of the algorithm, which
departs significantly from the existing 
analyzers for smart contracts~\cite{DBLP:conf/acsac/TorresSS18,mythril,oyente,manticore,
  DBLP:conf/ndss/KalraGDS18,DBLP:conf/isola/AltR18}, is to
automatically discover domain-specific invariants of smart contracts during the verification process. 
In particular, our algorithm automates the discovery of
{\em transaction invariants}, which are distinctive properties of
smart contracts that hold under arbitrary interleaving of
transactions and enable to analyze smart contracts exhaustively without exploring all  program paths separately. 
A technical challenge is to efficiently discover precise invariants from the huge search space.
We propose an effective algorithm tailored for typical smart contracts,
which iteratively generates and validates candidate invariants in a
feedback loop akin to the CEGIS (counter example-guided inductive
synthesis)
framework~\cite{Solar-Lezama:2006:CSF:1168917.1168907,Udupa:2013:TSP:2491956.2462174,Solar-Lezama:2008:PSS:1714168}.
Our algorithm is general and can be used for analyzing a wide range of
safety properties of smart contracts besides arithmetic safety.



Experimental results show that our algorithm is much more effective than existing techniques for analyzing Ethereum smart contracts. 
We first evaluated the
effectiveness of \verismart~by comparing it with four
state-of-the-art bug-finders: \osiris~\cite{DBLP:conf/acsac/TorresSS18},
\oyente~\cite{oyente}, \mythril~\cite{mythril}, and
\manticore~\cite{manticore}. An in-depth study on 60 
contracts that have CVE vulnerabilities shows that \verismart~detects all known 
vulnerabilities with a negligible false positive rate (0.41\%). By
contrast, existing bug-finders failed to detect a large amount ($>29.3\%$) of
known vulnerabilities with higher false positive rates ($>5.4\%$).
We also compared \verismart~with two state-of-the-art verifiers, 
\zeus~\cite{DBLP:conf/ndss/KalraGDS18} and
\smtchecker~\cite{DBLP:conf/isola/AltR18}. The results show that
\verismart~is significantly more precise than them thanks to its
 ability to discover transaction invariants of
smart contracts automatically. 


\myparagraph{Contributions}
Our contributions are as follows:
\begin{itemize}

\item We present a new verification algorithm for smart
  contracts (Section~\ref{sec:algorithm}). This is the first CEGIS-style algorithm that leverages transaction invariants automatically during the verification process.

\item We provide \verismart, a practical 
  implementation of our algorithm that supports the full Solidity
  language, the de facto standard programming language for writing Ethereum smart
  contracts.

    \item We provide in-depth evaluation of \verismart~in comparison with six
       analyzers~\cite{DBLP:conf/acsac/TorresSS18,oyente,mythril,manticore,DBLP:conf/ndss/KalraGDS18,DBLP:conf/isola/AltR18}. All experimental results are
      reproducible 
      as we make our tool and data publicly
      available.\footnote{\url{http://prl.korea.ac.kr/verismart}}
      




\end{itemize}

\section{Motivating Examples}
\label{sec:overview}

In this section, we illustrate central features of \verismart~with
examples. We use three real-world
smart contracts to highlight key aspects of \verismart~that differ from existing
 analyzers.




\myparagraph{Example 1}


Figure~\ref{fig:smt} shows a simplified function 
from the SmartMesh token contract (CVE-2018-10376).  In April 2018, an
attacker exploited a vulnerability in the function and succeeded to
create an extremely large amount of unauthorized tokens
($\approx 5\cdot 10^{57}$ USD).
This vulnerability, named proxyOverflow,
was due to unexpected 
integer
overflow.

The \texttt{transferProxy}
function 
is responsible for transferring a designated amount of tokens
(\texttt{value}) from a source address (\texttt{from}) to a
destination address (\texttt{to}) while paying transaction fees
(\texttt{fee}) to the message sender (\texttt{msg.sender}).  The core
functionality is implemented at lines 8--10, where the recipients'
balances (\texttt{balance[to]} and \texttt{balance[msg.sender]}) are
increased (lines 8 and 9) and the sender's balance
(\texttt{balance[from]}) is decreased by the same amount of the sent
tokens at line 10.


\begin{figure}
\begin{lstlisting}
function transferProxy (address from, address to, uint value, uint fee) {
  if (balance[from] < fee + value) revert();

  if (balance[to] + value < balance[to] ||
    balance[msg.sender] + fee < balance[msg.sender])
      revert();

  balance[to] += value;
  balance[msg.sender] += fee;
  balance[from] -= value + fee;
}
\end{lstlisting}
 \vspace{-0.5em}
\caption{A vulnerable function from SmartMesh (CVE-2018-10376).}
\label{fig:smt}
 \vspace{-0.5em}
\end{figure}

Note that the developer is aware of the risks of integer over/underflows and has
made effort to avoid them. 
The conditional statement at line 2 checks
whether the sender's balance ({\tt balance[from]}) is greater than or equal to the
tokens to be sent ({\tt fee+value}), aiming to prevent integer underflow at line
10.  The guard statements at lines 4 and 5 check that the
recipients' balances are valid after the transaction, intending to prevent
integer overflows at lines 8 and 9, respectively.

However, the contract still has a loophole at line 2. The expression {\tt fee+value} inside the
conditional statement may cause integer overflow, which enables the
token sender to send more money than (s)he has.
Suppose all accounts initially have no balances, i.e., 
$\texttt{balance[from]=0}$, $\texttt{balance[to]=0}$, and $\texttt{balance[msg.sender]=0}$,
and the function is invoked with the arguments
\texttt{value}=\texttt{0x8ff...ff} and 
\texttt{fee}=\texttt{0x700...01}, 
where 256-bit unsigned integer variables (\texttt{value} and
\texttt{fee}) are represented in hexadecimal numbers comprised of 64
digits (e.g., \texttt{value} has 63 \texttt{f}s and one \texttt{8}).
Suppose further the two unspecified address values are given as the same but
different from the sender's (i.e.,
$\texttt{from}=\texttt{to}\not=\texttt{msg.sender}$).  These crafted
inputs then make the sanity checks at lines 2--6 powerless
(i.e., the three conditions at lines 2, 4, and 5 are all false because
$\texttt{fee+value}=\texttt{0x8ff...ff}+\texttt{0x700...01} = \texttt{0}$ and
$\texttt{balance[to]} = \texttt{balance[msg.sender]} = \texttt{0}$). Therefore, lines
8--10 for token transfer are executed unexpectedly, 
creating a huge amount of tokens from nothing (i.e.,
$\texttt{balance[to]} = \texttt{balance[from]} = \texttt{0x8ff...ff}$ and
$\texttt{balance[msg.sender]} = \texttt{0x700...01}$.




This accident could have been prevented by \verismart, as it pinpoints
the vulnerability at line 2. 
Indeed, \verismart~is an exhaustive verifier,
aiming to detect all
arithmetic issues in smart contracts. By contrast, inexhaustive bug-finders
are likely to miss critical vulnerabilities. For example, 
among the
existing bug-finders~\cite{DBLP:conf/acsac/TorresSS18,oyente,mythril,manticore},
only \osiris~\cite{DBLP:conf/acsac/TorresSS18}~is able to find the
vulnerability. \mythril~\cite{mythril}~and \oyente~\cite{oyente} fail to detect the well-known proxyOverflow
vulnerability. 

\begin{figure}
  \begin{lstlisting}
function multipleTransfer(address[] to, uint value) { 
 require(value * to.length > 0);
 require(balances[msg.sender] >= value * to.length);
 balances[msg.sender] -= value * to.length; 
 for (uint i = 0; i < to.length; ++i) { 
   balances[to[i]] += value; 
 } 
}
\end{lstlisting}
 \vspace{-0.5em}
\caption{A vulnerable function from Neo Genesis Token (CVE-2018-14006).
}
\label{fig:multi}
 \vspace{-0.5em}
\end{figure}

\myparagraph{Example 2}
Figure~\ref{fig:multi} shows the {\tt multipleTransfer} function
adapted from the Neo Genesis Token contract (CVE-2018-14006).
%
The function has a similar vulnerability to that of the
first example.  At line 3 in Figure~\ref{fig:multi}, it
prevents the underflow possibility of the token sender's account but
does not protect the overflow of the tokens to be sent (\texttt{value
  * to.length}), which is analogous to the situation at line 2 of
Figure~\ref{fig:smt}.  That is, in a similar way, an attacker can send
huge amounts of tokens to any users by spending only few
tokens~\cite{cve14006-hack}.  

Despite the similarity between vulnerabilities in Example 1 and 2,
bug-finders have no guarantees of consistently finding them.
 For example, \osiris, which succeeded to detect the
vulnerability in Example 1, now fails to report the similar bug in
Example 2. The other bug-finders are ineffective too; \mythril~does 
not report any issues and \oyente~obscurely reports that the entire
function body is vulnerable without specifying certain operations. 
On the other hand, \verismart~reliably reports that the expression {\tt value *
  to.length} at lines 2--4 would overflow.

One of the main reasons for the unstable results of bug-finders is
that they rely heavily on a range of heuristics to avoid false
positives (e.g., see~\cite{DBLP:conf/acsac/TorresSS18}). 
Though  heuristics are good at reducing false positives, the resulting analyzer
is often very brittle; even small changes in programs may end up with
missing fatal vulnerabilities as shown in Example 1 and 2, which is
particularly undesirable for safety-critical software like smart contracts.

\myparagraph{Example 3}

Figure~\ref{fig:cve13326} shows a simplified version of the contract,
called BTX.  The program has two global state variables:
\texttt{balance} stores balances of each account address (line 2), and
\texttt{totalSupply} is the total amount of the supplied tokens (line
3).  The constructor function initializes \texttt{totalSupply} with
$10000$ tokens (line 6), and gives the same amount of tokens to the
creator of the contract (line 7).  The \texttt{transfer} function
sends \texttt{value} tokens from the transaction message sender's
account to the recipient's account (lines 12--13), if it does not
incur the underflow in the message sender's balance (line 11).  The
\texttt{transferFrom} function is similar to \texttt{transfer} with an
exception to the order of performing addition and subtraction.

The contract has four arithmetic operations at lines 12, 13, 18, and
19, all of which are free of integer over/underflows.  However, it
is nontrivial to see why they are all safe.  In particular, the safety
of the two addition operations at lines 13 and 18 is tricky, because
there are no direct safety-checking statements in each function.  To
see why they do not overflow, we need to discover the following two
{\em transaction invariants} that always hold no
matter how the transactions ({\tt transfer} and {\tt transferFrom})
are interleaved:
\begin{itemize}
\item the sum of all account values is $10000$, i.e.,
  \begin{equation}\label{eq:inv1}
    \sum_{i}\mbox{\tt balance[$i$]} = 10000,
  \end{equation}
\item and computing $\sum_{i}\mbox{\tt balance[$i$]}$ does not cause
  overflow.
\end{itemize}
By combining these two conditions and the preconditions expressed in
the {\tt require} statements at lines 11 and 17,
we can conclude that, at lines 13 and 18, the maximum values of both
\texttt{balance[to]} and \texttt{value} are $10000$, and
thus the expression \texttt{balance[to]+value} does not overflow in
256-bit unsigned integer operations.

Since reasoning about the safety in this case is tricky, it is likely for human
auditors to make a wrong conclusion that the contract is unsafe. This
is in fact what happened in the recent CVE report (CVE-2018-13326)\footnote{\url{https://nvd.nist.gov/vuln/detail/CVE-2018-13326}};
the CVE report incorrectly states that the two addition operations at
lines 13 and 18 are vulnerable
and thus the operations may overflow.
Unfortunately, existing safety analyzers do not help here.
In particular, verifiers, \zeus~\cite{DBLP:conf/ndss/KalraGDS18}
and \smtchecker~\cite{DBLP:conf/isola/AltR18}, are not precise enough
to keep track of the implicit invariants such as
(\ref{eq:inv1}) and therefore cannot prove the safety at lines 13 and 18. 
Bug-finders \osiris~and
\oyente~also produce false alarms. \mythril~does not report
any issues, but this does not mean that
it proved the absence of vulnerabilities. 

By contrast, \verismart~is able to 
prove that the contract is
safe without any false alarms. Notably, \verismart~does so
by automatically inferring hidden invariants described above. 
To our knowledge, \verismart~is the first of its kind,
which discovers global invariants of smart
contracts and leverages them during the verification process in a
fully automated way. 

\begin{figure}
\begin{lstlisting}
contract BTX {
  mapping (address => uint) public balance;
  uint public totalSupply;

  constructor () {
    totalSupply = 10000;
    balance[msg.sender] = 10000;
  }

  function transfer (address to, uint value) {
    require (balance[msg.sender] >= value);
    balance[msg.sender] -= value;
    balance[to] += value; // Safe
  }

  function transferFrom (address from, address to, uint value) {
    require (balance[from] >= value);
    balance[to] += value; // Safe
    balance[from] -= value;
  }
}
\end{lstlisting}
 \vspace{-0.5em}
\caption{Example contract simplified from CVE-2018-13326.
}
\label{fig:cve13326}
 \vspace{-0.5em}
\end{figure}




\section{\verismart~Algorithm}\label{sec:algorithm}

This section describes the verification algorithm of \verismart. 
We formally present the algorithm in a general
setting, so it can be used for analyzing other safety properties as
well beyond our application to arithmetic safety.

\myparagraph{Language}
For brevity, we focus on a core subset of Solidity~\cite{solidity}.  However, \verismart~supports the full Solidity language as the extension is discussed in Section~\ref{sec:implementation}.
Consider the following subset of Solidity: 
\[
\begin{array}{l}
c \in C ~::=~ G^*~ F^*, \qquad f \in F ~::=~ x(y) \{ S \}\\
a \in A ~::=~ x:= E \mid x[y]:=E \mid \assumecmd(B) \mid \assertcmd(B) \\
s \in S~::=~ A \mid   \ifcmd~B~S_1~S_2  \mid \whilecmd^l ~E~S \mid S_1;S_2 \\
\end{array}
\]

We assume a single contract $c$ is given, which consists of a sequence of global state variable
declarations ($G^*$) and a sequence of function definitions ($F^*$),
where $G$ and $F$ denote the sets of global variables and functions in
the contract, respectively.  We assume a constructor function
$f_0 \in F$ exists in $c$.  Each function $f$ is defined by a function
name ($x$), argument ($y$), and a body statement ($S$).
A statement $S$ is an atomic statement ($A$), a conditional statement,
or a while loop. An atomic statement $a \in A$ is an assignment to a variable ($x:=E$), an assignment
to an array element ($x[y]:=E$), an $\assumecmd$ statement, or an
$\assertcmd$ statement. In our language, we model mapping variables in
Solidity as arrays.
In our language, $\assumecmd$
differs from $\assertcmd$; while the former models the {\tt require}
statements in Solidity and stops execution if the condition evaluates
to false, the latter does not affect program semantics.
$E$ and $B$ stand for conventional arithmetic
and boolean expressions, respectively, where we assume
arithmetic expressions produce 256-bit unsigned integers. 
In our language, loops are annotated with labels ($l$), and
the entry and the exit of each function $f$ are annotated with special labels
$\lentry_f$ and $\lexit_f$, respectively. Let $\mylabel$ be the set of
all labels in the program. 
We assume each function $f$ has \texttt{public} (or \texttt{external})
visibility, meaning that all functions in the contract can be called
from the outside. 

\myparagraph{Goal} 
Our goal is to develop an algorithm that proves or disproves
every assertion (which we also call {\em query}) in the contract.  We
assume that safety properties to verify are expressed as 
the $\assertcmd$ statements in the program. 
In our application to arithmetic safety, assertions can
be automatically generated; for example, 
for each addition \texttt{a+b} and multiplication \texttt{a*b}, 
we generate \texttt{assert(a+b>=a)} and
\texttt{assert(a==0||(a!=0 \&\& (a*b)/a==b))}, respectively.  

\myparagraph{Notation} 
We use the lambda notation for functions. For
example, $\lambda x. x+1$ is the function that takes $x$
and returns $x+1$. 
We write $\fol$ for the set of first-order formulas
in the combined theory of  fixed-sized bitvectors, arrays with extensionality, and equality with uninterpreted functions.
When $e$ is an expression or a formula, we write $e[y/x]$ for the new
expression where $x$ gets replaced by $y$. 
We write $\free(e)$ for the set of free variables in $e$.  

\subsection{Algorithm Overview}\label{sec:alg-overview}

\verismart~departs significantly from existing analyzers for smart contracts~\cite{DBLP:conf/acsac/TorresSS18,mythril,oyente,manticore,
  DBLP:conf/ndss/KalraGDS18,DBLP:conf/isola/AltR18,
  Nikolic:2018:FGP:3274694.3274743, DBLP:conf/ccs/TsankovDDGBV18,
  Grech:2018:MSO:3288538.3276486, vandal} in that \verismart~applies a CEGIS-style verification algorithm that 
iteratively searches for hidden invariants that are required for verifying safety properties. 

\myparagraph{Invariants of Smart Contracts}

We consider two kinds of invariants for smart contracts: transaction and loop invariants.  We say a formula is a
transaction invariant if it is valid at the end of the constructor and
the validity is preserved by the execution of public functions that
can be invoked by transactions. Loop
invariants are more standard; a formula is an invariant of a loop if the formula is valid at the
entry of the loop and is preserved by the loop body.  Transaction
invariant is global and thus it is a single formula, whereas loop
invariants are local and must be separately given for each loop in the
program. Thus, our algorithm aims to discover a pair $(\ti,
\inducmap)$, where 
$\ti \in \fol$ is a transaction invariant and
$\inducmap \in \mylabel \to \fol$ is a mapping from loop labels to
formulas.
We write $\bigwedge$ for pointwise conjoining
operation between two mappings $\inducmap_1$ and $\inducmap_2$, i.e.,
$\inducmap_1 \bigwedge \inducmap_2 = \lambda l \in \mylabel. \inducmap_1(l) \land \inducmap_2(l)$. 

\begin{figure}
\begin{lstlisting}
contract RunningExample {
  uint public n;
  constructor () { n = 1;}
  function f () public {
    assert (n + 1 >= n);
    n = n + 1;
    if (n >= 100) { n = 1; }
  }
}
\end{lstlisting}
 \vspace{-0.5em}
\caption{Example contract.}
\label{fig:simple}
 \vspace{-0.5em}
\end{figure}

\medskip
\begin{example}
Consider the contract in Figure~\ref{fig:simple}. 
The program has one global variable \texttt{n}, which is initialized
to \texttt{1} in the constructor.
The function \texttt{f} can be invoked from the outside of the contract;
it increases the value of \texttt{n} by 1 every time it is called,
but resets it to \texttt{1} whenever \texttt{n} is \texttt{100}.
Note that $n \le 100$ is a
transaction invariant: 1) it holds at the end of
the constructor, and 2) supposing that $n \le 100$ holds before
entering \texttt{f}, we can prove that it also holds
when exiting the function. 
Our algorithm automatically discovers the invariant $n \le 100$ and
succeeds to prove that the assertion at line 5 is safe; upon entering \texttt{f}, $n \le 100$ holds
and $n \le 100 \to n + 1 \ge n$ is valid in the theory of unsigned 256
bitvector arithmetic.
\end{example}

\begin{figure}[t]
\center
\includegraphics[scale=0.25]{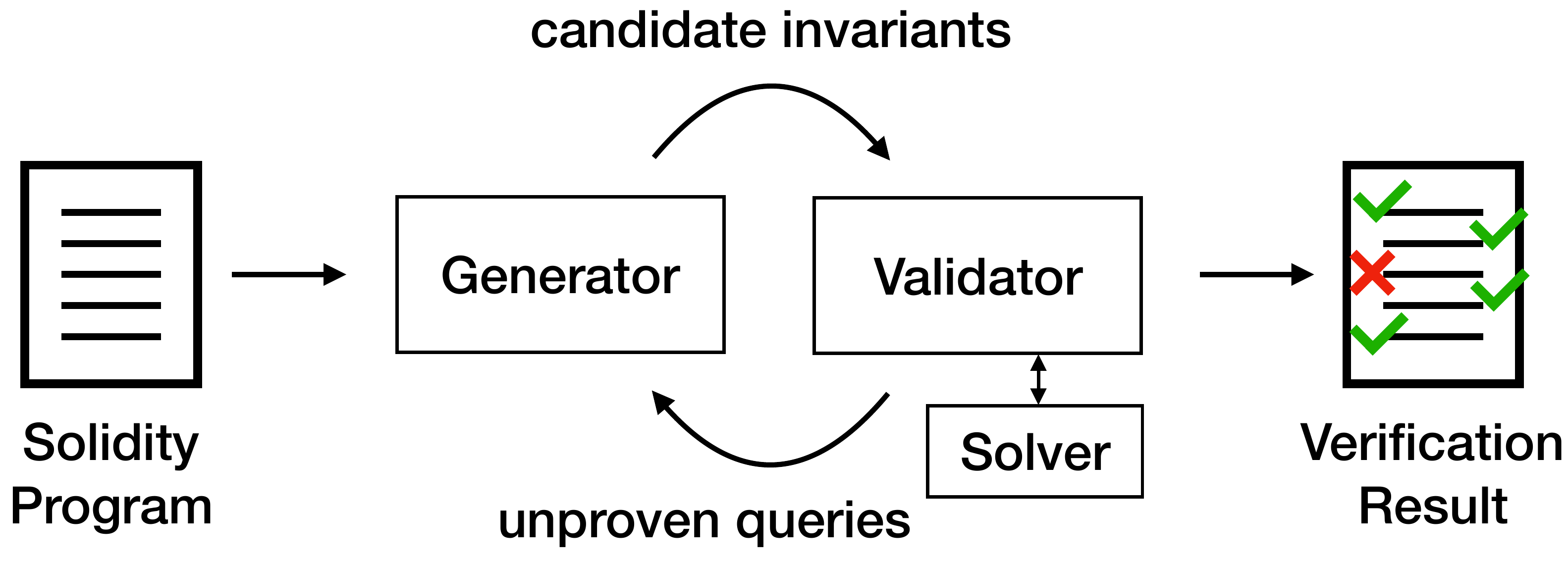}
\caption{Algorithm~overview.}
\label{fig:overview}
 \vspace{-1em}
\end{figure}

\myparagraph{Algorithm Structure}
Figure~\ref{fig:overview} describes the overall structure of our algorithm. 
The input is a smart contract written in Solidity, and the output is
a verification result that indicates whether each query (i.e.,
assertion) in the program is proven safe or not. 
The algorithm consists of two components, a validator
and a generator, where the validator has a solver as a subcomponent. 

The algorithm aims to find contract-specific invariants that are
inductive and strong enough to prove all provable queries in the given
contract. The role of the generator is to produce candidate invariants
that help the validator to prove as many queries as possible. Given a
candidate invariant, the validator checks whether the invariant is
useful for proving the queries. If it fails to prove the queries, it
provides the set of unproven queries as feedback to the 
generator. The generator uses this feedback to refine the current
invariant and generate new ones.  This way, the validator and
generator form an iterative loop that continuously refines the
analysis results until the program is proven to be safe or the given
time budget is exhausted. Upon termination, all unproven queries are
reported to users as potential safety violations.





Algorithm~\ref{algpseudo} shows our verification algorithm.
It uses a workset ($W$) to maintain candidate invariants,
which initially contains the trivial invariant
$(\true, \lambda l.\true)$ (line 1): the transaction invariant $\ti$
is $\true$ and the loop invariant mapping $\inducmap$ maps every label ($l$)
to $\true$. The repeat-until loop at lines
2--11 correspond to the feedback loop in
Figure~\ref{fig:overview}.  At lines 3 and 4, the algorithm chooses
and removes a candidate invariant $(\ti, \inducmap)$ from the
workset. 
We choose a candidate invariant that is the smallest in size. 
At line 5, we run the validator to check whether the current
candidate is inductive and strong enough to prove queries, which
returns a pair of the boolean variable ${\it inductive}$, indicating
whether the current candidate invariant is inductive or not,
and the set $U$ of unproven queries.
If $U$ is empty (line 6), the
algorithm terminates and the contract is completely proven to be
safe. Otherwise (line 8), we generate a new set of candidate
invariants and add them to the workset.
Finally, when the current candidate fails to prove some
queries but is known to be at least inductive (line 9),
we strengthen the remaining candidate invariants using it (line 10),
because we can potentially prove more queries with stronger invariants.
By doing so, we can find useful invariants more efficiently.
The algorithm iterates until it times out or the workset becomes empty.
We assume that the algorithm implicitly maintains previously
generated invariants to avoid redundant trials. 

\myparagraph{Technical Contributions}
Although the overall algorithm follows the general framework of CEGIS~\cite{Solar-Lezama:2006:CSF:1168917.1168907,Udupa:2013:TSP:2491956.2462174,Solar-Lezama:2008:PSS:1714168}, we provide an effective, domain-specific instantiation of the framework in the context of smart contract analysis. 
Now we describe the details of this instantiation: validator (\ref{sec:vc}), generator (\ref{sec:generator}), and solver (\ref{sec:solver}). 


\begin{algorithm}[t]
  \caption{Our Verification Algorithm}
  \label{algpseudo}
  \begin{algorithmic}[1]
    \Require A smart contract $c$ to verify
    \Ensure Verification success or potential safety violations
    \State $W \gets \myset{ (\true, \lambda l. \true)}$
    \Repeat
    \State Choose a candidate invariant $(\ti, \inducmap)$ from $W$
    \State $W \gets W \setminus \myset{(\ti, \inducmap)}$
    \State $ ({\it inductive}, U)  \gets \Verify (c, \ti, \inducmap)$
    \If {$U = \emptyset$} {verification succeeds}
    \Else
    \State $W \gets W \cup \UpdateInv (U, \ti, \inducmap)$    
     \If {$\it inductive$}
     \State  $W \gets \myset{(\ti' \land \ti, \inducmap' \bigwedge \inducmap) \mid (\ti',\inducmap') \in W}$ \EndIf
    \EndIf
    \Until{$W = \emptyset$ or timeout} 
    \State {\bf return}~potential safety violations
  \end{algorithmic}
\end{algorithm}


\subsection{Validator}
\label{sec:vc}

The goal of the validator is to check whether the current candidate
invariant $(\ti, \inducmap)$ is inductive and strong enough to prove
 safety of the queries.  The input to the validator is an {\em
  annotated program} $(c, \ti, \inducmap)$, i.e., smart contract
$c$ annotated with transaction ($\ti$) and loop ($\inducmap$)
invariants. The validator proceeds in three steps.

\myparagraph{Basic Path Construction} 
Given an annotated
program $(c, \ti, \inducmap)$, we first break down the program into a
finite set of basic paths~\cite{Bradley:2007:CCD:1324777}.
A basic path is a sequence of
atomic statements that begins at the entry of a function or a loop,
and ends at the exit of a function or the entry of a loop, without
passing through other loop entries.
We represent a basic path $p$ by the five components: $((l_1,\phi_1), a_1;\dots; a_n, (l_2,\phi_2))$, 
where $l_1$ is the label of the starting point (i.e., function or loop entry) of the path, $\phi_1
\in\fol$ is the invariant annotated at $l_1$, $a_1,\dots,a_n$ are
atomic statements, $l_2$ is the label of the end point (i.e., function exit or loop entry) of the path,  and
$\phi_2 \in \fol$ is the invariant annotated at $l_2$.
The basic path satisfies the following properties:
\begin{enumerate}
\item If $l_1$ is a function entry, $\phi_1 = \ti$ (i.e., transaction invariant). An exception: $\phi_1 = \true$ if
  $l_1$ is entry of constructor. 
 If $l_2$ is  a function exit,
  $\phi_2 = \ti$. 
\item Otherwise, i.e., when $l_1$ and $l_2$ are labels of
  loops, $\phi_1 = \inducmap(l_1)$ and $\phi_2 = \inducmap(l_2)$ (i.e., considering loop invariants).
\end{enumerate}
Note that our construction of basic paths is exhaustive as we 
consider {\em all} paths of the program by summarizing the effects of transactions and loops   with their invariants. The basic paths can be computed  by traversing control flows of the program. 

\begin{example}\label{ex:basicpaths}
\label{ex:bp}
Consider the contract in Figure~\ref{fig:simple} annotated with the transaction
invariant $\ti = n \le 100$. We do not consider loop invariants as the
contract does not have any loops. The annotated program is converted
into three basic paths:
\[
\begin{array}{l}
p_1: ((\lentry_0, \true) , n:=1 , (\lexit_0, n \le 100)) \\
p_2: ((\lentry_f, n \le 100), a_1, (\lexit_f, n \le 100)) \\
p_3: ((\lentry_f, n \le 100), a_2, (\lexit_f, n \le 100)) \\
\end{array}
\]
where $a_1 = \assertcmd(n+1\ge n); n:=n+1; 
 \assumecmd(n\ge 100);n:=1$ and $a_2 = \assertcmd(n+1\ge n); n:=n+1;
 \assumecmd(n<  100)$. 
$p_1$ represents the basic path of the constructor (whose entry and
exit labels are $\lentry_0$ and $\lexit_0$, respectively). 
$p_2$ and $p_3$ represent the basic paths of the function {\tt f} that
follow the true and false branches of the conditional statement at
line 7, respectively. Note that conditional statements and loops do not appear
as they are broken into basic paths with original conditions given as $\assumecmd$ statements. 
\end{example}

\myparagraph{Generation of Verification Conditions}
Let $P$ be the set of basic paths constructed from the annotated program.
We next generate verification conditions (VCs) for each basic
path. 

To derive the VCs, we should be able to express
effects of program statements in $\mathsf{FOL}$.
To do so, we define a strongest postcondition
predicate transformer $\post: \mathsf{stmt} \to 
\fol\times \fol \to \fol \times \fol$,
which is defined in a standard way for each atomic statement as follows: 
\[\small
\begin{array}{r@{\;\;}c@{\;\;}l}
\post(x:=e)(\phi_1, \phi_2) &= & (x=e[x'/x] \land \phi_1[x'/x], \phi_2) \\
\post(x[y]:=e) (\phi_1, \phi_2) &=& (x=x' \langle y \vartriangleleft
                                    e[x'/x]\rangle \land \phi_1[x'/x], \phi_2)  \\
\post(\assumecmd(e))(\phi_1, \phi_2) &=& (\phi_1 \land e, \phi_2) \\
\post(\assertcmd(e))(\phi_1, \phi_2) &=& (\phi_1, \phi_2 \land (\phi_1 \to e)) \\
\end{array}
\]
where unprimed variables (e.g., $x$) and primed variables (e.g., $x'$) represent the current and previous program states, respectively. 
In each rule, $\phi_1$ is a precondition and $\post$ transforms it into
a postcondition while accumulating the safety conditions of assertions
in $\phi_2$. 
We write $x' \langle y \vartriangleleft e \rangle$ for the
modified array $x'$ that stores the value of $e$ at position $y$.
With $\post$, we define the procedure $\genvc$ that generates the
VC of a basic path: 
\[
  \genvc (((l_1,\phi_1), a_1;\dots; a_n, (l_2,\phi_2))) =
  ( \phi'_1 \to \phi_2,  \phi'_2)
\]
where $(\phi'_1, \phi'_2) = (\post(a_n) \circ  \dots \circ \post(a_2)
\circ \post(a_1)) (\phi_1, \true)$.
The generated VC consists of two parts:
$\phi'_1 \to \phi_2$ is a formula for checking that the annotated
invariants are inductive, and $\phi'_2$ is a formula for checking the
safety properties in assertions.


\begin{example}\label{ex:vc-example}
Consider the basic path $p_3$ in Example~\ref{ex:basicpaths}.
The corresponding VC is a pair of
$(n' \le 100 \land n=n'+1 \land n < 100 \to n \le 100, n \le 100 \to n + 1 \ge n)$,
both of which are valid in the bitvector theory.
\end{example}

 \myparagraph{Collecting Unproven Paths}
Finally, we return a pair of the boolean variable ${\it inductive}$ and
the subset $U \subseteq P$ of basic paths whose VCs are
invalid:
\[
\begin{array}{l}
({\it inductive} , U) =  \\
\quad \left \{
\begin{array}{l}
\mbox{\bf if}~\exists p \in P. \genvc(p).1~\mbox{is invalid}~\mbox{\bf then}  \\
\qquad  (\false, \myset{p \in P \mid \genvc(p).1~\mbox{is invalid}})  \\
\mbox{\bf else}~(\true, \myset{p \in P \mid \exists F \in \genvc(p).2~\mbox{is invalid}} ) 
\end{array}
\right.
\end{array}
\]
$\genvc(p).1$ and $\genvc(p).2$ denote the first (i.e., the VC on inductiveness) and
the second (i.e., the VC on safety) component of $\genvc(p)$, respectively.
We also write $F \in \genvc(p).2$ for a clause
of $\genvc(p).2$, where $F$ corresponds to the safety
condition of a single query.  
In the above procedure, we first check
whether some VCs regarding inductiveness are invalid.
If it does so (if-case),
we set ${\it inductive}$ to $\false$ and 
$U$ becomes the basic paths where inductiveness checking failed.
Note that, in this case, we accelerate our verification procedure by excluding from $U$
the paths where safety checking may fail. 
That is, we first focus on refining invariants to be inductive and then strengthen them further to prove safety rather than trying to achieve both at the same time. 
When the current candidate invariant is inductive (else-case), we set ${\it
  inductive}$ to $\true$ and collect the basic paths 
where some queries are not proven to be safe.
To check the validity of the VCs, we use a
domain-specific solver, which will be explained in
Section~\ref{sec:solver}.
   






\subsection{Generator} \label{sec:generator}

The generator takes the set $U$ as feedback and
produces new candidate invariants by refining the current one
$(\ti, \inducmap)$.
$\UpdateInv(U, \ti, \inducmap)$ returns the following set: 
\begin{align*}
  \myset{(\ti, \inducmap') \mid 
                                   \inducmap' \in \RefineLoop(\inducmap, U)} 
                                  \cup
  \myset{(\ti', \inducmap) \mid \ti' \in \RefineTran (\ti, U)}
\end{align*}
where $\RefineLoop$ and $\RefineTran$ generate
new loop and transaction
invariants, respectively, based on the current
ones. We define $\RefineLoop(\inducmap, U)$ so as to return the
following set of refined loop invariants:
\[
\begin{small}
  \bigcup_{((l_1,\_), a, (l_2, \_)) \in U}
  \myset{ \inducmap[
    l_i \mapsto \phi_i ] \mid i \in [1,2], \phi_i \in 
     \nextloopinv(\inducmap(l_i), a) } 
      \end{small}
 \]
where we assume $l_1$ and $l_2$ are loop labels, and $a$ is the
  sequence of atomic statements in the basic path. The definition of $\RefineTran(\ti, U)$:
  \[
    \myset{\ti' \mid ((l_1,\_), a, (l_2, \_)) \in U, \ti' \in \nexttraninv(\ti,
    a)}
  \]
where we assume $l_1$ is the label of a function entry or
      $l_2$ is the label of a function exit. 
In the definitions above, the procedures
      $\nextloopinv$ and $\nexttraninv$ are actually responsible for 
      refining loop and transaction invariants, which ultimately
      determine the effectiveness of the generator and  the overall
      verification algorithm.






\myparagraph{Domain-Specific Refinement}

We define $\nextloopinv$ and $\nexttraninv$ in terms of 
{\em refinement relation}.
A refinement relation $(\mytrans_{X,C}) \subseteq \fol \times \fol$ is
a binary relation on logical formulas, parameterized by variable set $X$ and constant set
$C$, which describes how a candidate invariant is refined in one step:
i.e., $\phi$ can be refined to any of
$\myset{\phi' \mid \phi \mytrans_{X,C} \phi'}$. 
In our approach, choosing a right refinement relation
holds the key to cost-effective verification since it 
defines the search space of candidate invariants. 
For example, simply choosing a very general or specific refinement relation would not be
practical because of the huge or too limited search space. 
Instead, we have to carefully design a 
refinement relation tailored for real-world smart contracts to make
our algorithm cost-effective.

Fortunately, we observed that smart contracts in practice
share common properties and accordingly considered the following points when we
design the refinement relation. 
First, smart contracts often use loops in simple and restricted
    forms, e.g.,  {\tt for(i = 0; i < x ; i++)}, and therefore it is sufficient to consider simple numerical
invariants. In particular, we decided to focus on invariants of the forms $x =
    y$, $x \ge y$, $x = n$, $x \ge n$, and $x \le n$, where $x, y$ are
    variables and $n$ denotes integer constants. That is, we do not consider
    non-linear or compound invariants such as $x = y^2$ and $x = y +
    z$.
Second, because smart contracts use the mapping datatype extensively
    (e.g., {\tt balance} in token contracts), it is particularly
    important to capture their common properties (e.g., the sum of {\tt
      balance} is equal to {\tt totalSupply}). Currently, we support
    the function symbol $\mysigma$ for variables of mapping type: for
    example,    $\mysigma$$(${\tt balance}$)$ means the sum of all
    balances. 
Third, we consider invariants that are quantifier-free conjunctive
formulas. That is, we do not allow disjunctions or quantifiers to be
used in candidate invariants. 
  

Based on the observations, we define the refinement relation: 
\[
\begin{array}{c}
\phi_1 \mytrans_{X,C} \phi_2 \iff \phi_2 = \phi_1 \land
  \varphi~\mbox{and}~\varphi \in A
\end{array}
\]
where $A$ is the set of atomic predicates of the forms 
$x=y, x \ge y, x=n, x \ge n, x \le n, \mysigma(x) = e$, where $x, y
\in X$, $n \in C$, and $e \in C \cup X$.
That is, the current invariant $\phi_1$ is strengthened with a linear
and quantifier-free atomic predicate ($\varphi$). Note that we only
use the symbol $\mysigma$ in the equality predicate as we found invariants
of other forms such as $\mysigma(x) > e$ are rarely used in practice.
Finally, we define $\nexttraninv$ and $\nextloopinv$ using $\mytrans_{X,C}$ as follows: 
\[
\begin{array}{rcl}
\nextloopinv(\ti, a) &=& \myset{ \ti' \mid \ti \mytrans_{{\it vars}(a)
                        , {\it const}(a)} \ti'} \\
\nexttraninv(\phi, a) &=& \myset{ \phi' \mid \phi \mytrans_{{\it
                          globals}, {\it cnstr} \cup
                          {\it const}(a)} \phi'} \\
\end{array}
\]
where ${\it vars(a)}$ and ${\it const(a)}$ are the variables and
constants appearing in the atomic statements $a$, respectively. {\it globals}
and {\it cnstr} represent the set of global variables and constants in
the constructor function, respectively.  We instantiate the sets $X$
and $C$ differently because transaction invariants often involve
global state variables and constants of the entire contract while loop
invariants involve local and global variables and constants that
appear in the enclosing function. In both cases, we reduce
the search space by focusing on local variables and constants to those
 of  the current basic path ($a$).

\subsection{Solver}\label{sec:solver}

The last component is the solver that is used by
the validator to discharge the verification conditions. The solver ultimately
uses an off-the-shelf SMT solver (we use Z3~\cite{z3paper}) 
but performs domain-specific preprocessing and optimization steps before using it, which we found important to make our approach practical for real-world contracts.
For a basic path $p$, we assume its verification condition
$F$ (either the inductiveness condition, i.e., $F=\genvc(p).1$, or
the safety condition of a query, i.e., $F \in \genvc(p).2$) is given.

\myparagraph{Preprocessing}

Since $F$ may contain symbols (i.e., $\mysigma$) that conventional SMT
solvers cannot understand, we must preprocess $F$ so that all such
uninterpretable symbols get replaced by equi-satisfiable formulas in 
conventional theories.
For example, let $F$ contains $\mysigma$ as follows:
\[
F = \cdots \land \mysigma(x) = n \land x[i] = v_1 \land x[j] = v_2 \land  \cdots
\]
where we elide portions of $F$ that are irrelevant to the mapping
variable $x$ (i.e., $x$ is only accessed with $i$ and $j$ in the given basic path $p$). 
Our idea to translate $F$ into a formula without $\mysigma$ is
to instantiate the symbol with respect to the context where
$F$ is evaluated. In this example, we can translate the formula $F$ into
the following:
\[
\cdots \land  F_1 \land F_2 \land x[i] = v_1 \land x[j] = v_2 \land \cdots
\]
where $F_1 = (i \not= j \to x[i] + x[j] + R_x = n) \land 
(i = j \to x[i] + R_x = n)$ 
asserts that the sum of distinct elements of $x$ equals 
$n$. Because $x$ is used in the given basic path with two index
variables $i$ and $j$, we consider two cases: $i=j$ and $i\not=j$.
When $i\not=j$, we replace $\mysigma(x)=n$ by $x[i] + x[j] + R_x = n$,
where $R_x$ is a fresh variable denoting the sum of $x[k]$ for all $k
\in {\it domain}(x) \setminus \myset{i,j}$, where ${\it domain}(x)$ is
the domain of the mapping. The other case ($i=j$) is
handled similarly.  $F_2$ is the additional assertion that guarantees
the validity of $F_1$: $F_2 = (i \not= j \to x[i] + x[j] \ge x[j] \land 
x[i] + x[j] + R_x \ge R_x) \land 
(i = j \to x[i] + R_x \ge R_x) \land B_x$, where $B_x$ is a fresh
propositional variable, meaning that the summations in $R_x$ do not
overflow.
The general method for our preprocessing is given in Appendix~\ref{sec:sum}.

Note that the verification condition after preprocessing can be
checked by a conventional SMT solver. However, we found that the
resulting formulas are often too complex for modern SMT solvers to
handle efficiently, so we apply the following optimization techniques.
%

\myparagraph{Efficient Invalidity Checking}\label{sec:invalidity}
Most importantly, we quickly decide invalidity of formulas without invoking SMT solvers. 
We observed that even state-of-the-art SMT solvers can be extremely inefficient
when our verification conditions are invalid. 
For example, consider the following formula: 
\[
\true \to (a-b=0) \lor (a-b\not=0 \land ((a-b)*255) / (a-b) = 255).
\]
It is easy to see that the formula  is invalid in the theory of
256-bit arithmetic (e.g., it does not hold when $a=2^{255}$ and $b=0$).
Unfortunately, however, the latest version of Z3~\cite{z3paper} (ver 4.8.4) and CVC4~\cite{BCD+11} (ver 1.7)
takes more than 3 minutes to conclude the formula is invalid.

To mitigate this problem, we designed a simple decision procedure 
based on the free variables of formulas;
given a VC of the form $p \to q$, we conclude that it is invalid if 
$\free (p) \not\supseteq \free(q)$. 
The intuition is that $p$ must include more
variables than $q$, as a necessary condition to
be stronger than $q$. In the above example,
we conclude the formula is invalid because
$\free(\true) \not\supseteq \free (a=0 \lor (a\not=0 \land (a * b) / a = b)) = \myset{a,b}$. In practice, we found that this simple technique improves the scalability of the verification algorithm significantly as it avoids expensive calls to SMT solvers. 

Let us explain why our technique is correct. 
We first review the notion of interpretation in first-order logic~\cite{Bradley:2007:CCD:1324777}.
An interpretation $I: (D_I, \alpha_I)$ is a pair of a domain ($D_I$) and an assignment ($\alpha_I$). The domain $D_I$ is a nonempty set of
  values (or objects). The assignment $\alpha_I$ maps variables, constants, functions,
  and predicate symbols to elements, functions, and predicates over $D_I$.
  Let $J: I \vartriangleleft \myset{x \mapsto v}$ denote
  an $x$-variant of $I$ such that $J$ accords with $I$ on everything except for $x$. That is, $D_I = D_J$ and $\alpha_I[y]=\alpha_J[y]$ if $y \not= x$, but
  $\alpha_I[x]$ and $\alpha_J[x]$ may be different.
Then, we have the following result (see Appendix~\ref{sec:invalidity-checking} for proof).
\begin{prop}\label{prop:invalidity-checking}
Let $p$ and $q$ be first-order formulas.  
Then, $p \to q$ is invalid if the following three conditions hold:
\begin{enumerate}[(i)]
\item\label{sitm1} $\free(p) \not\supseteq \free(q)$, 
\item\label{sitm2} $p$ is satisfiable: $\exists I.\; I \models p$, and
\item\label{sitm3} $q$ has a nontrivial variable: there exists $x \in \free(q) \setminus \free(p)$ such that 
for any interpretation $I$, if $I \models q$ then $I \vartriangleleft\myset{x \mapsto v}
    \models \neg q$ for some $v \in D_I \setminus \myset{\alpha_I[x]}$.
\end{enumerate}
\end{prop}
Our technique is based on this result but checks the first condition (\ref{sitm1}) only, which can be done syntactically and efficiently. We do not check the last two conditions (\ref{sitm2}) and (\ref{sitm3}) as they require invoking SMT solvers in general. Therefore, our technique may decide valid VCs as invalid (i.e., producing false positives) although no invalid VCs are determined to be valid (i.e., no false negatives). Because the technique causes no false negatives, it can be used by sound verifiers. 

Although approximated, our technique rarely produces false positives in practice. 
For example, consider the valid formula $\true \to a \ge a$. Our technique may incorrectly conclude that the formula is invalid,
since $\free(\true) \not\supseteq \free(a \ge a)$ but we do not check the condition (\ref{sitm3}) that the formula violates.
Note that, however, such a {\em trivial} formula is unlikely to appear during the verification of real-world smart contracts; the verification condition $\true \to a \ge a$ would be generated from the trivial expression $a-a$ that does not appear frequently in programs. Even when they appear, we can easily remove the {\em triviality}. For example, it is easy to simplify $\true \to a \ge a$ into $\true \to \true$ that is not determined as invalid by our technique since $\free(\true) \supseteq \free(\true)$. In fact, 
no false positives were caused by our technique in our experiments in Section~\ref{sec:evaluation}.

\myparagraph{Efficient Validity Checking}\label{sec:template}
We also quickly identify some valid formulas by using a number of domain-specific templates. This is because our verification conditions are likely to involve 
arrays and non-linear expressions extensively but 
modern SMT solvers are particularly inefficient for handling them.
For example, a simple yet important validity template is as follows:
\[
\infer[n_1 \le n_2]
{F' \to x \ge (x*n_1) /n_2}
{}
\]
where $F'$ denotes an arbitrary formula, $x$ a 256-bit unsigned integer variable,
and $n_1$ and $n_2$ some integer constants.
This template asserts that, regardless of the precondition $F'$,
$x \ge (x*n_1) /n_2$  holds
if $n_1 \le n_2$.
Using the template, we can conclude that a formula
$\dots \to y \ge (y * 99) / 100$ is valid (i.e., the subtraction $y - (y * 99) / 100$ is
safe from underflow)
without calling an external SMT solver.
These templates are used before the preprocessing step;
several templates were designed to determine the validity of formulas containing
domain-specific symbols at a high level without preprocessing.
We provide more examples in Appendix~\ref{sec:template-example}.






\section{Implementation}\label{sec:implementation}

In this section, we explain implementation details of \verismart,
which consists of about 7,000 lines of OCaml code.
Although Section~\ref{sec:algorithm} describes our algorithm
for a small subset of Solidity, our implementation supports
the full language (except for inline assembly).
Most Solidity features (e.g., function modifers) can be desugared into our core language in a
straightforward way. 
We discuss nontrivial issues below.

\myparagraph{Function Calls}
Basically, we handle function calls by inlining them into their call-sites up to a
predefined inlining depth $k$ (currently, less than or equal to 2). Exceptions include
relatively large functions (with more than 20 statements) that might cause scalability
issues and inter-contract function calls (i.e., calling
functions in other contracts via contract objects).
To perform exhaustive verification, we handle those
remaining function calls conservatively as follows.


First, we conservatively reflect side-effects of function calls on the caller side.
To do so, we first run a  side-effect analysis~\cite{Cooper:1988:ISA:53990.53996} to find variables
whose values may be changed by the called functions. Next, we weaken the
formulas at call-sites by replacing each of atomic predicates that
involve those variables by \true. 
For example, consider a call statement \texttt{x:=foo()} and assume
\texttt{foo} may change the value of variable \texttt{a} in its body.
Suppose further the precondition of the call-site is $a \ge 1 \land b \ge 1 \land c \ge 1 \land x \ge y$.
Then, we obtain the following postcondition of the call-site:
$\true \land b \ge 1 \land c \ge 1 \land \true$ where
$a \ge 1$ and $x \ge y$ get replaced by $\true$. 
Regarding inter-contract function calls, it is enough to invalidate
the value of return variables only, as inter-contract calls in
Solidity cannot directly modify other contracts' states. 
For example, consider the precondition above and an inter-contract
call \texttt{x : = o.foo ()}. We produce the postcondition $a \ge 1 \land b \ge 1 \land c \ge 1 \land \true$,
where only $x \ge y$ is replaced by $\true$. 


Second, we separately analyze function bodies not 
inlined. This step is needed to
detect potential bugs in the functions skipped during the step
described in the preceding paragraph.
To perform exhaustive verification, we analyze these functions by
over-approximating their input states.
Specifically, when the function in a main contract has
\texttt{public} or \texttt{external} visibility, we run the
algorithm in Section~\ref{sec:algorithm} which annotates entry and
exit with transaction invariant.
On the other hand, when the function in a main contract
has \texttt{internal} or \texttt{private} visibility
(i.e., the functions which cannot be called from the outside
and can only be accessed via function call statements)
or the function is defined in other contracts,
we generate the VCs after we annotate entries and exits of them with \true,
i.e., incoming state at the entry is over-approximated as \true~and
inductiveness condition can be trivially checked at the exit.


In summary, \verismart~performs exhaustive safety verification
without missing any possible behaviors.
In theory, we may lose precision due to the conservative function-call
analysis. However, as our experimental results in
Section~\ref{sec:evaluation} demonstrate, our approach
is precise enough  in practice. 




\myparagraph{Inheritance}
In Section~\ref{sec:algorithm}, we assumed a single contract is given. 
To support contract inheritance,
we copy functions and global variables of 
parent contracts to a main contract using the inheritance graph
provided by the Solidity compiler. During this conversion, we consider
function overriding and variable hiding, and do not copy functions with the same signatures
and the same variables.

\myparagraph{Structures}
We encode structures in Solidity with arrays.
To do so, we introduce a special mapping variable for each member of a
structure type, which maps structures to the member values.
For example, given a precondition $\phi$, the strongest postcondition
of command \texttt{x.y := z} is $m_y = m'_y \langle x  \vartriangleleft z \rangle \land \phi[m'_y / m_y]$,
where $m_y$ is a map (or an array) from structures to the
corresponding values of member \texttt{y} and
$x$ is an uninterpreted symbol for the structure variable \texttt{x}.
Note that we are able to handle aliasing among structures
using this encoding.
For example,
if two structures \texttt{p} and \texttt{q} are aliased and
they both have \texttt{y} as a member, then
we can access the same member \texttt{y} using
either of the structures, i.e., $m_y[p]=m_y[q]$.

\myparagraph{Inline Assembly}
One potential source of false negatives of source code analyzer (e.g.,
\zeus~\cite{DBLP:conf/ndss/KalraGDS18}) is inline assembly.
\verismart~also has this limitation and may miss bugs hidden in
embedded bytecode.
However, \verismart~conservatively analyzes the remaining parts of the source code
by considering the side-effects of the assembly blocks in a similar way that
we handle function call statements, i.e.,
we replace each atomic predicate by \true~if it involves variables
used in assembly code (using the information provided by the
Solidity compiler). 
Note that this limitation does not impair the practicality of
\verismart~significantly, as inline assembly is not very common in
practice. For example, 
in our benchmarks in Section~\ref{sec:evaluation}, only four contracts (\#4, \#16, \#52 in Table~\ref{table:main},
\#24 in Table~\ref{table:zeus}) contain assembly blocks but none of
these assembly blocks include arithmetic operations. 

\section{Evaluation}
\label{sec:evaluation}


\setlength{\dashlinedash}{2.0pt}
\setlength{\dashlinegap}{1.0pt}

We evaluate 
the effectiveness of \verismart~by
comparing it with existing tools. Research questions are as follows:
\begin{enumerate}[(1)]
\item How precisely can \verismart~detect arithmetic bugs compared to
  the existing bug-finders, i.e., \osiris~\cite{DBLP:conf/acsac/TorresSS18}, \oyente~\cite{oyente},
  \mythril~\cite{mythril}, \manticore~\cite{manticore}?
  \item How does \verismart~compare to the existing verifiers, i.e., \zeus~\cite{DBLP:conf/ndss/KalraGDS18} and \smtchecker~\cite{DBLP:conf/isola/AltR18}?
  \end{enumerate}
In addition, we conduct a case study to show 
\verismart~can be easily extended to support other
types of vulnerabilities (Section~\ref{sec:checker}).
We used the latest versions of the existing tools (as of May 1st, 2019).
All experiments were conducted on a machine with
Intel Core i7-9700K and 64GB RAM.


\subsection{Comparison with Bug-finders}
\label{sec:precision}

We evaluate the bug-finding capability of \verismart~by comparing it
with four bug-finding analyzers for Ethereum smart contracts:
\osiris~\cite{DBLP:conf/acsac/TorresSS18}, \oyente~\cite{oyentetool},
\mythril~\cite{mythril}, and \manticore~\cite{manticore}.
They are well-known open-sourced tools that
support detection of integer overflows
(\osiris, \oyente, \mythril, \manticore) and division-by-zeros (\mythril).
In particular, \osiris~is arguably the state-of-the-art tailored for finding
integer overflow bugs~\cite{DBLP:conf/acsac/TorresSS18}.

\myparagraph{Setup}
We used 60 smart contracts that
have vulnerabilities with assigned CVE IDs. We have chosen these
contracts to enable in-depth manual study on the analysis results
with known vulnerabilities confirmed by CVE reports.
The 60 benchmark contracts were selected randomly from the
487 CVE reports that are related to arithmetic overflows (Table~\ref{table:cve}),
excluding duplicated contracts with minor syntactic
differences 
(e.g., differences in contract names or logging events).
During evaluation, we found four incorrect CVE reports
(\#13, \#20, \#31, \#32 in Table~\ref{table:main}), which
will be discussed in more detail at the end of the section.

To run \osiris, \oyente, \mythril, and \manticore, we
used public docker images provided together with these tools.
Following prior work~\cite{DBLP:conf/acsac/TorresSS18}, we set the
timeout to 30 minutes per contract. For fair comparison, we activated only the analysis
modules for arithmetic bug detection when such option is available (\mythril, \manticore). 
We left other options as
default.
For \verismart, we set the timeout to 1 minute for the last
entrance of the loop in Algorithm~\ref{algpseudo},
and set the timeout to 10 seconds for Z3 request,
because these numbers worked effectively in our experience;
if we set each timeout to a lower value,
the precision may decrease (Section~\ref{sec:threat}).
In analysis reports of each tool, we only counted alarms related to
arithmetic bugs (integer over/underflows and division-by-zeros)
for a main contract whose name is available at the Etherscan website~\cite{etherscan}.

\myparagraph{Results}
\renewcommand{\arraystretch}{0.85}
\begin{table*}[t]
\caption{
Evaluation of existing tools on CVE reports.
\textsf{LOC}: lines of code.
\Q: the total number of queries for each contract after removing unreachable functions.
\textsf{\#Alarm}: the number of entire alarms produced by each tool.
\FP: the number of false alarms.
\CVE: a marker that indicates whether each tool successfully detects vulnerabilities in CVE.
\cmark: a tool successfully pinpoints all vulnerable locations in CVE.
  $\triangle$: a tool detects only a part of vulnerabilities in CVE, or obscurely reports that an entire function body is vulnerable
  without pinpointing specific locations.
  \xmark: a tool totally failed to detect vulnerabilities in CVE.
N/A: all vulnerabilities reported in CVE are actually safe
(\#13, \#31).
For partly correct CVE reports (\#20, \#32), the \CVE~information is valid w.r.t. them. 
}
\begin{scriptsize}
\setlength{\tabcolsep}{.45em}
\begin{tabular}{|l clrr| rrc| rrc| rrc| rrc| rrc|}
\hline
\multirow{2}{*}{\textsf{No.}} & \multirow{2}{*}{\textsf{CVE ID}} & \multirow{2}{*}{\textsf{Name}} & \multicolumn{1}{c}{\multirow{2}{*}{\textsf{LOC}}} & \multicolumn{1}{c}{\multirow{2}{*}{\Q}} &
\multicolumn{3}{|c|}{\verismart}  & \multicolumn{3}{c|}{\osiris~\cite{DBLP:conf/acsac/TorresSS18}}& \multicolumn{3}{c|}{\oyente~\cite{oyente,oyentetool}}& \multicolumn{3}{c|}{\mythril~\cite{mythril}}    & \multicolumn{3}{c|}{\manticore~\cite{manticore}}  \\
\cline{6-20}
 &&   & \multicolumn{1}{c}{} & \multicolumn{1}{c|}{} & \multicolumn{1}{c}{\Alarms} & \multicolumn{1}{c}{\FP} & \CVE & \multicolumn{1}{c}{\Alarms} & \multicolumn{1}{c}{\FP} & \CVE & \multicolumn{1}{c}{\Alarms} & \multicolumn{1}{c}{\FP} & \CVE & \multicolumn{1}{c}{\Alarms} & \multicolumn{1}{c}{\FP} & \CVE & \multicolumn{1}{c}{\Alarms} & \multicolumn{1}{c}{\FP} & \CVE \\
 \hline
\#1  & 2018-10299 & BEC                & 299 & 6  & 2  & 0 & \cmark & 0  & 0 & \xmark & 1  & 0 & $\triangle$ & 2     & 0 & \cmark & 0                       & 0 & \xmark \\
\#2  & 2018-10376 & SMT                & 294 & 22 & 13 & 0 & \cmark & 1  & 0 & \cmark & 2  & 0 & \xmark & 1     & 0 & \xmark & \multicolumn{3}{c|}{timeout ($>$ 3 days)}   \\
\#3  & 2018-10468 & UET                & 146 & 27 & 14 & 0 & \cmark & 9  & 0 & \xmark & 8  & 0 & \cmark & 5     & 0 & \cmark & 0                       & 0 & \xmark \\
\#4  & 2018-10706 & SCA                & 404 & 48 & 33 & 0 & \cmark & 9  & 0 & \xmark & 4  & 0 & $\triangle$ & 2     & 0 & \xmark & \multicolumn{3}{c|}{internal error}     \\
\#5  & 2018-11239 & HXG                & 102 & 11 & 7  & 0 & \cmark & 6  & 0 & \cmark & 2  & 0 & \xmark & 3     & 0 & \cmark & 2                       & 0 & \cmark \\
\#6  & 2018-11411 & DimonCoin          & 126 & 15 & 7  & 0 & \cmark & 5  & 0 & \xmark & 5  & 0 & \cmark & 5     & 0 & \cmark & 3                       & 0 & \cmark \\
\#7  & 2018-11429 & ATL                & 165 & 9  & 4  & 0 & \cmark & 3  & 0 & \cmark & 2  & 0 & $\triangle$ & 0     & 0 & \xmark & 0                       & 0 & \xmark \\
\#8  & 2018-11446 & GRX                & 434 & 39 & 24 & 2 & \cmark & 8  & 2 & \xmark & 12 & 4 & \xmark & 4     & 2 & \xmark &  \multicolumn{3}{c|}{internal error}     \\
\#9  & 2018-11561 & EETHER             & 146 & 10 & 5  & 0 & \cmark & 4  & 0 & \cmark & 2  & 0 & $\triangle$ & 2     & 0 & \cmark & 0                       & 0 & \xmark \\
\#10 & 2018-11687 & BTCR               & 99  & 20 & 4  & 0 & \cmark & 2  & 0 & \cmark & 2  & 0 & $\triangle$ & 3     & 2 & \xmark & 0                       & 0 & \xmark \\
\#11 & 2018-12070 & SEC                & 269 & 40 & 8  & 0 & \cmark & 6  & 0 & \cmark & 4  & 0 & \xmark & 3     & 1 & \xmark & 0                       & 0 & \xmark \\
\#12 & 2018-12230 & RMC                & 161 & 9  & 5  & 0 & \cmark & 3  & 0 & \cmark & 5  & 0 & \cmark & 0     & 0 & \xmark & 0                       & 0 & \xmark \\
\#13 & 2018-13113 & ETT                & 142 & 9  & 2  & 0 & N/A & 4  & 2 & N/A & 2  & 2 & N/A & 0     & 0 & N/A & 0                       & 0 & N/A \\
\#14 & 2018-13126 & MoxyOnePresale     & 301 & 5  & 3  & 0 & \cmark & 0  & 0 & \xmark & 0  & 0 & \xmark & 0     & 0 & \xmark & 0                       & 0 & \xmark \\
\#15 & 2018-13127 & DSPX               & 238 & 6  & 4  & 0 & \cmark & 3  & 0 & \cmark & 3  & 0 & $\triangle$ & 1     & 0 & \xmark & 0                       & 0 & \xmark \\
\#16 & 2018-13128 & ETY                & 193 & 10 & 4  & 0 & \cmark & 3  & 0 & \cmark & 3  & 0 & $\triangle$ & 0     & 0 & \xmark & 0                       & 0 & \xmark \\
\#17 & 2018-13129 & SPX                & 276 & 9  & 6  & 0 & \cmark & 5  & 0 & \cmark & 3  & 0 & $\triangle$ & 1     & 0 & \xmark &  \multicolumn{3}{c|}{internal error}   \\
\#18 & 2018-13131 & SpadePreSale       & 312 & 4  & 3  & 0 & \cmark & 0  & 0 & \xmark & 0  & 0 & \xmark & 0     & 0 & \xmark & \multicolumn{3}{c|}{internal error}   \\
\#19 & 2018-13132 & SpadeIco           & 403 & 9  & 6  & 0 & \cmark & 0  & 0 & \xmark & 0  & 0 & \xmark & 0     & 0 & \xmark & \multicolumn{3}{c|}{internal error}     \\
\#20 & 2018-13144 & PDX                & 103 & 5  & 2  & 0 & \cmark & 2  & 1 & \cmark & 2  & 1 & \cmark & \multicolumn{3}{c|}{internal error}    & 0                       & 0 & \xmark \\
\#21 & 2018-13189 & UNLB               & 335 & 4  & 3  & 0 & \cmark & 2  & 0 & \cmark & 3  & 0 & \cmark & 1     & 0 & \xmark & 0                       & 0 & \xmark \\
\#22 & 2018-13202 & MyBO               & 183 & 17 & 11 & 0 & \cmark & 5  & 0 & \cmark & 3  & 0 & \xmark & 1     & 0 & \xmark &  \multicolumn{3}{c|}{internal error}   \\
\#23 & 2018-13208 & MoneyTree          & 171 & 17 & 10 & 0 & \cmark & 4  & 0 & \cmark & 2  & 0 & \xmark & 2     & 0 & \xmark & 0                       & 0 & \xmark \\
\#24 & 2018-13220 & MAVCash            & 171 & 15 & 10 & 0 & \cmark & 4  & 0 & \cmark & 2  & 0 & \xmark & 1     & 0 & \xmark & 0                       & 0 & \xmark \\
\#25 & 2018-13221 & XT                 & 186 & 15 & 10 & 0 & \cmark & 4  & 0 & \cmark & 2  & 0 & \xmark & 2     & 0 & \xmark & 0                       & 0 & \xmark \\
\#26 & 2018-13225 & MyYLCToken       & 181	& 17  & 11   & 0 & \cmark 	&  5    & 0  & \cmark  &   6    & 0   & 	\xmark	&    0    & 0  & \xmark &   0   & 0 & \xmark 		 \\
\#27 & 2018-13227 & MCN                & 172 & 17 & 10 & 0 & \cmark & 4  & 0 & \cmark & 2  & 0 & \xmark & 2     & 0 & \xmark & 0                       & 0 & \xmark \\
\#28 & 2018-13228 & CNX                & 171 & 17 & 10 & 0 & \cmark & 4  & 0 & \cmark & 2  & 0 & \xmark & 2     & 0 & \xmark & 0                       & 0 & \xmark \\
\#29 & 2018-13230 & DSN                & 171 & 17 & 10 & 0 & \cmark & 4  & 0 & \cmark & 2  & 0 & \xmark & 2     & 0 & \xmark & 0                       & 0 & \xmark \\
\#30 & 2018-13325 & GROW               & 176 & 12 & 2  & 0 & \cmark & 4  & 2 & \cmark & 1  & 1 & \xmark & 0     & 0 & \xmark & 0                       & 0 & \xmark \\
\#31 & 2018-13326 & BTX                & 135 & 9  & 2  & 0 & N/A & 4  & 2 & N/A & 2  & 2 & N/A & 0     & 0 & N/A & 0                       & 0 & N/A \\
\#32 & 2018-13327 & CCLAG              & 92  & 5  & 2  & 0 & \cmark & 2  & 1 & \cmark & 2  & 1 & \cmark & 0     & 0 & \xmark & 0                       & 0 & \xmark \\
\#33 & 2018-13493 & DaddyToken         & 344 & 40 & 22 & 0 & \cmark & 8  & 0 & \xmark & 2  & 0 & \xmark & 3     & 0 & \xmark &  \multicolumn{3}{c|}{internal error}     \\
\#34 & 2018-13533 & ALUXToken          & 191 & 23 & 13 & 0 & \cmark & 8  & 0 & \cmark & 2  & 0 & \cmark & 1     & 0 & \xmark & 1                       & 0 & \xmark \\
\#35 & 2018-13625 & Krown              & 271 & 22 & 9 & 0 & \cmark & 1  & 0 & \xmark & 3  & 0 & \cmark & 0     & 0 & \xmark & \multicolumn{3}{c|}{internal error}    \\
\#36 & 2018-13670 & GFCB               & 103 & 14 & 11 & 0 & \cmark & 6  & 1 & \cmark & 3  & 1 & \cmark & 1     & 0 & \xmark & 0                       & 0 & \xmark \\
\#37 & 2018-13695 & CTest7             & 301 & 17 & 8  & 0 & \cmark & 0  & 0 & \xmark & 0  & 0 & \xmark & 0     & 0 & \xmark & 0                       & 0 & \xmark \\
\#38 & 2018-13698 & Play2LivePromo     & 131 & 8  & 7  & 0 & \cmark & 7  & 0 & \cmark & 7  & 0 & \cmark & 5     & 0 & \xmark & 5                       & 0 & \xmark \\
\#39 & 2018-13703 & CERB\_Coin         & 262 & 17 & 8  & 0 & \cmark & 5  & 0 & \cmark & 2  & 0 & \xmark & 2     & 1 & \xmark & 0                       & 0 & \xmark \\
\#40 & 2018-13722 & HYIPToken          & 410 & 8  & 3  & 0 & \cmark & 2  & 0 & \cmark & 2  & 0 & \cmark & 0     & 0 & \xmark &  \multicolumn{3}{c|}{internal error}     \\
\#41 & 2018-13777 & RRToken            & 166 & 8  & 3  & 0 & \cmark & 2  & 0 & \cmark & 2  & 0 & \cmark & 0     & 0 & \xmark & 0                       & 0 & \xmark \\
\#42 & 2018-13778 & CGCToken           & 224 & 13 & 6  & 0 & \cmark & 4  & 0 & \cmark & 4  & 0 & \cmark & 1     & 0 & \xmark & 1                       & 0 & \xmark \\
\#43 & 2018-13779 & YLCToken           & 180 & 17 & 11 & 0 & \cmark & 5  & 0 & \cmark & 6  & 0 & \cmark & 0     & 0 & \xmark & 0                       & 0 & \xmark  \\
\#44 & 2018-13782 & ENTR               & 171 & 17 & 10 & 0 & \cmark & 4  & 0 & \cmark & 2  & 0 & \cmark & 2     & 0 & \xmark & 0                       & 0 & \xmark \\
\#45 & 2018-13783 & JiucaiToken        & 271 & 19 & 11 & 0 & \cmark & 6  & 0 & \cmark & 4  & 0 & \cmark & 0     & 0 & \xmark &  \multicolumn{3}{c|}{internal error}    \\
\#46 & 2018-13836 & XRC                & 119 & 22 & 7  & 0 & \cmark & 5  & 0 & \xmark & 3  & 0 & $\triangle$ & 3     & 1 & \cmark & \multicolumn{3}{c|}{timeout ($>$ 3 days)}  \\
\#47 & 2018-14001 & SKT                & 152 & 19 & 10 & 0 & \cmark & 4  & 0 & \xmark & 3  & 0 & $\triangle$ & 3     & 0 & \cmark & 0                       & 0 & \xmark \\
\#48 & 2018-14002 & MP3                & 83  & 12 & 4  & 0 & \cmark & 2  & 0 & \xmark & 2  & 0 & $\triangle$ & 2     & 1 & \xmark & \multicolumn{3}{c|}{timeout ($>$ 3 days)}    \\
\#49 & 2018-14003 & WMC                & 200 & 15 & 6  & 0 & \cmark & 3  & 0 & \xmark & 2  & 0 & $\triangle$ & 3     & 0 & \cmark & 1                       & 0 & \xmark \\
\#50 & 2018-14004 & GLB                & 299 & 40 & 8  & 0 & \cmark & 5  & 0 & \cmark & 1  & 0 & $\triangle$ & 0     & 0 & \xmark & 0                       & 0 & \xmark \\
\#51 & 2018-14005 & Xmc                & 255 & 29 & 11 & 0 & \cmark & 8  & 0 & \cmark & 1  & 0 & $\triangle$ & 3     & 0 & $\triangle$ & 0                       & 0 & \xmark \\
\#52 & 2018-14006 & NGT                & 249 & 27 & 13 & 0 & \cmark & 1  & 0 & \xmark & 5  & 0 & $\triangle$ & 0     & 0 & \xmark &\multicolumn{3}{c|}{timeout ($>$ 3 days)}    \\
\#53 & 2018-14063 & TRCT               & 178 & 9  & 1  & 0 & \cmark & 1  & 0 & \cmark & 1  & 0 & \cmark & 4     & 2 & \cmark & 0                       & 0 & \xmark \\
\#54 & 2018-14084 & MKCB               & 273 & 17 & 10 & 0 & \cmark & 5  & 0 & \cmark & 4  & 0 & \xmark & 2     & 0 & \xmark & 1                       & 0 & \xmark \\
\#55 & 2018-14086 & SCO                & 107 & 16 & 14 & 0 & \cmark & 7  & 2 & \cmark & 5  & 2 & \xmark & 0     & 0 & \xmark & 0                       & 0 & \xmark \\
\#56 & 2018-14087 & EUC                & 174 & 15 & 7  & 0 & \cmark & 4  & 0 & \xmark & 4  & 0 & \xmark & 0     & 0 & \xmark & 0                       & 0 & \xmark \\
\#57 & 2018-14089 & Virgo\_ZodiacToken & 208 & 30 & 20 & 0 & \cmark & 12 & 0 & \cmark & 5  & 0 & \cmark & 14    & 0 & \cmark & 0                       & 0 & \xmark \\
\#58 & 2018-14576 & SunContract        & 194 & 12 & 4  & 0 & \cmark & 1  & 0 & \cmark & 0  & 0 & \xmark & 0     & 0 & \xmark & 0                       & 0 & \xmark \\
\#59 & 2018-17050 & AI                 & 141 & 8  & 3  & 0 & \cmark & 1  & 0 & \cmark & 1  & 0 & \cmark & 0     & 0 & \xmark & 0                       & 0 & \xmark \\
\#60 & 2018-18665 & NXX                & 79  & 7  & 5  & 0 & \cmark & 4  & 0 & \cmark & 4  & 0 & \cmark & 0     & 0 & \xmark & 0                       & 0 & \xmark \\
\hline
\multicolumn{3}{|c}{\multirow{3}{*}{\textbf{Total}}} & & && & \multicolumn{1}{r|}{\cmark:58} & &     & \multicolumn{1}{r|}{\cmark:41}& & & \multicolumn{1}{r|}{\cmark:20}& && \multicolumn{1}{r|}{\cmark:10}&& & \multicolumn{1}{r|}{\cmark:{ }2}   \\
\multicolumn{3}{|l}{} &  12493& 976& 492   & 2&  \multicolumn{1}{r|}{$\triangle$:{ }0} & 240&13& \multicolumn{1}{r|}{$\triangle$:{ }0} 	& 171 & 14 & \multicolumn{1}{r|}{$\triangle$:15}  &  94  	& 10      & \multicolumn{1}{r|}{$\triangle$:{ }1}   & 14    &0    & \multicolumn{1}{l|}{$\triangle$: 0}    \\
\multicolumn{3}{|l}{} 	& &  	&  & 	& \multicolumn{1}{r|}{\xmark{ }:{ }0}& & &  \multicolumn{1}{r|}{\xmark{ }:17}&   &  & \multicolumn{1}{r|}{\xmark{ }:23}  &&       &  \multicolumn{1}{r|}{\xmark{ }:46}   &    &    & \multicolumn{1}{r|}{\xmark{ }:42}  \\
\hline
\end{tabular}
\end{scriptsize}
\label{table:main}
\vspace{-0.5em}
\end{table*}

Table~\ref{table:main} shows the evaluation results on the CVE
dataset. For each benchmark contract and tool, the table shows the
number of alarms (\Alarms) and the number of false
positives (\FP) reported by the tool;
regarding these two numbers,
we did not count cases where the tools (\oyente~and \mythril)
ambiguously report that the entire body of a function or
the entire contract is vulnerable.
The \CVE~columns indicate whether
the tool detected the vulnerabilities in CVE reports or not
(\cmark: a tool successfully pinpoints all
vulnerable locations in each CVE report,
\xmark: a tool does not detect any of them,
$\triangle$: a tool detects only a part of
vulnerable points in each CVE report or,
obscurely reports the body of an entire function containing
CVE vulnerabilities is vulnerable
without pinpointing specific locations.
N/A: all vulnerabilities in CVE reports are actually safe;
see Table~\ref{table:cve-incorrect}).

The results show that \verismart~far outperforms the existing bug-finders
in both precision and recall.  In total, \verismart~reported 492 arithmetic over/underflow and
division-by-zero alarms. We carefully inspected these alarms and
confirmed that 490 out of 492 were true positives (i.e., safety can be
violated for some feasible inputs), resulting in a false
positive rate (${\FP}\over{\Alarms}$) of 0.41\% (2/492).
We also inspected 484 (=976-492) unreported
queries to confirm that all of them are true negatives (i.e., no
feasible inputs exist to violate safety), resulting in a recall of
100\%. Of course, \verismart~detected all CVE vulnerabilities.
In contrast, existing bug-finders missed many vulnerabilities.
For example, \osiris~managed to detect 41 CVE vulnerabilities with 17
undetected known vulnerabilities.
\oyente~ pinpointed 20 exact vulnerable locations in CVE,
partly detected vulnerabilities in 4 CVE reports,
vaguely raised alarms on 11 functions containing
vulnerable locations, and missed 23 CVE vulnerabilities.
\mythril~detected vulnerabilities in 10 CVE reports,
obscurely warned that 1 function is vulnerable,
and missed 46 known issues. 
\manticore~was successful in only two CVE reports,
failing on 42 CVE reports.
The false positive rates of \osiris, \oyente, and \mythril~were 5.42\%
(13/240), 8.19\% (14/171), and 10.64\% (10/94), respectively.

\myparagraph{Efficiency}
\verismart~was also competitive
in terms of efficiency. 
To obtain the results in Table~\ref{table:main} on the 60 benchmark programs,
\verismart, \osiris, \oyente, \mythril, and \manticore~took
1.1 hour (3,807 seconds), 4.2 hours (14,942 seconds),
14 minutes, 
13.8 hours (49,680 seconds), and 31.4 hours
(112,920 seconds) respectively,
excluding the cases of timeout
(though we set the timeout to 30 minutes, \manticore~sometimes did not terminate within 3 days)
and internal errors (e.g., unsupported operations encountered,
abnormal termination) of \mythril~and \manticore.

\myparagraph{False Alarms of Bug-finders}

To see why \verismart~achieves higher precision than bug-finders, we inspected all 37 (=13+14+10) false positives
reported by bug-finders. 
Bug-finders reported 18 among 37 false positives due to the lack of inferring transaction invariants, all of which are avoided by \verismart. 
The remaining 19 false positives were due to imprecise handling of conditional statements. 
For example, consider the following code snippet (from \#55):
\begin{lstlisting}[numbers=none]
function transfer(address _to, uint _value) {
  if (msg.sender.balance < min)
    sell((min - msg.sender.balance) / sellPrice);
}
\end{lstlisting}
where the safety of \texttt{min - msg.sender.balance} is ensured by
the preceding guard. 
Both \osiris~and~\oyente~incorrectly reported that the subtraction is unsafe and integer underflow would occur. This might be because \osiris~and \oyente~do not keep track of complex path conditions (e.g., involving structures in this case) for some engineering issues. 
In contrast, \verismart~analyzes every conditional statement precisely
and do not produce such false alarms.

\myparagraph{False Alarms of \verismart}
\verismart~produced two false alarms in the benchmark \#8, because
it is currently unable to capture quantified transaction invariants.
Consider the \texttt{unlockReward} function in Figure~\ref{fig:FP}.
The subtraction operation at line 5 seems to cause arithmetic underflow;
the \texttt{value} may be changed at line 4, and thereafter the relation
\texttt{totalLocked[addr] > value} seems not to hold anymore.
However, the subtraction is safe
because the following transaction invariant holds over the entire
contract:
  \begin{equation}\label{eq:fpinv}
    \forall \mbox{\tt x}. \mbox{\tt totalLocked[x]} = \sum_{i}\mbox{\tt locked[x][$i$]}
  \end{equation}
with an additional condition that computing the summation ($\sum_{i}\mbox{\tt
  locked[x][$i$]}$) does not cause overflow.
With this transaction invariant, \texttt{value} is always less than \texttt{totalLocked[addr]}.
Because \verismart~considers quantifier-free invariants only (Section~\ref{sec:generator}), it falsely
reported that an underflow would occur at line 5.
\osiris~and \oyente~produced the false alarm too at the same location.

\begin{figure}
\begin{lstlisting}
function unlockReward(address addr, uint value) {
  require(totalLocked[addr] > value);
  require(locked[addr][msg.sender] >= value);
  if(value == 0) value = locked[addr][msg.sender];
  totalLocked[addr] -= value;  // false positive
  locked[addr][msg.sender] -= value;
}
\end{lstlisting}
\caption{
A function simplified from the benchmark \#8.
\osiris, \oyente, and \verismart~warn that
the subtraction at line 5 can cause arithmetic underflow, which is
false positive (i.e., the subtraction is safe).
}
\label{fig:FP}
\vspace{-0.5em}
\end{figure}

\myparagraph{False Negatives of Bug-finders}
We inspected CVE vulnerabilities that were commonly
missed by the four bug-finders, and
we found that the bug-finders often fail to detect bugs when
vulnerabilities could happen via inter-contract
function calls.  
For example, consider  code adapted from \#18:
\begin{lstlisting}[numbers=none]
function mint (address holder, uint value) {
  require (total+ value <= TOKEN_LIMIT); // CVE bug
  balances[holder] += value; 		         // CVE bug
  total += value;									  	 	 // CVE bug
}
\end{lstlisting}
\vspace{-0.5em}
There is a function call \texttt{token.mint (...,...)} in a main contract, where  \texttt{token} is a contract object. We can  see that
all three addition operations possibly overflow with some inputs.
For example, suppose \texttt{total=1}, $\texttt{value=0xfff\dots ff}$,
and $\texttt{TOKEN\_LIMIT=10000}$. Then, \texttt{total+value} overflows in
unsigned 256-bit and thus the safety checking statement can be bypassed.
Next, if \texttt{balances[holder]=0}, the \texttt{holder}
can have tokens more than the predetermined
limit \texttt{TOKEN\_LIMIT}. \verismart~detected the bugs as it conservatively analyzes inter-contract calls (Section~\ref{sec:implementation}).



\myparagraph{Incorrect CVE Reports Found by \verismart}
Interestingly, \verismart~unexpectedly identified six incorrectly-reported
CVE vulnerabilities. 
In Table~\ref{table:cve-incorrect},
the column \textsf{\# Incorrect Queries} 
denotes the number of queries incorrectly reported to be vulnerable for each CVE ID. 
We could discover them as \verismart~did not produce any alarms for those queries and then we manually confirmed that the CVE reports are actually incorrect. 
We have submitted a request for revising these issues to the CVE assignment team.

With the capability of automatically computing transaction invariants,
\verismart~successfully proved the safety for all the incorrectly
reported vulnerabilities (i.e., zero false positives). In other words, \verismart~could not have discovered incorrect CVE reports if it were without transaction invariants. 
The transaction invariants generated for proving the safety were similar to those in
Example 3 of Section~\ref{sec:overview}.
In contrast, existing bug-finders cannot be used for this purpose
such as proving the safety;
for example, \osiris~and \oyente~produced false positives
for \textit{all} of the 6 safe queries (i.e., the 6 incorrectly reported queries).

\label{sec:incorrect-cve}
\begin{table}[t]
\caption{
List of incorrect CVE reports found by \verismart.
\textsf{\#Incorrect Queries}: the number of incorrectly reported queries to be vulnerable.
\FP: the number of alarms raised by each tool for the incorrectly
reported queries.
}
\label{table:cve-incorrect}
\setlength{\tabcolsep}{.38em}
\center
\begin{tabular}{|c|l|r|r|r|r|}
\hline
\multirow{2}{*}{\textsf{CVE ID}} & \multirow{2}{*}{\textsf{Name}} & \multicolumn{1}{c|}{\textsf{\#Incorrect}}   & \multicolumn{3}{c|}{\FP}     \\ \cline{4-6}
                        &                       & \multicolumn{1}{c|}{\textsf{Queries}} & \multicolumn{1}{c|}{\osiris} & \multicolumn{1}{c|}{\oyente} & \multicolumn{1}{c|}{\verismart} \\ \hline
2018-13113              & ETT                   & 2                     & 2                           & 2                           & 0                            \\
2018-13144              & PDX                   & 1                    & 1                           & 1                           & 0                            \\
2018-13326              & BTX                   & 2                     & 2                           & 2                           & 0                            \\
2018-13327              & CCLAG                 & 1                 & 1                           & 1                           & 0                            \\ \hline
\end{tabular}
\vspace{-0.5em}
\end{table}



\subsection{Comparison with Verifiers}
We now compare \verismart~with  \smtchecker~\cite{DBLP:conf/isola/AltR18} and \zeus~\cite{DBLP:conf/ndss/KalraGDS18},
two recently-developed verifiers for smart contracts.
In particular, \smtchecker~is the ``official'' verifier for Ethereum smart contracts 
developed by the
Ethereum Foundation, which is available in the Solidity compiler. 
Like \verismart, the
primary goal of \smtchecker~is to detect arithmetic over/underflows
and division-by-zeros~\cite{DBLP:conf/isola/AltR18}. 



\myparagraph{Setup}
First of all, we must admit that the comparison with \zeus~and
\smtchecker~in this subsection is rather limited, because \zeus~is not
publicly available
and \smtchecker~is currently an experimental tool
that does not support the full Solidity language.
Since we cannot run \zeus~on our dataset, the only option was to use the public evaluation
data~\cite{zeus-report} provided by the \zeus~authors.
However, the public data was not detailed enough to accurately
interprete as the \zeus~authors classify each benchmark contract
simply as `safe' or `unsafe' without specific alarm information such
as line numbers.
The only objective information we could obtain from the data~\cite{zeus-report} was the fact that
\zeus~produces some (nonzero) number of false (arithmetic-overflow) alarms on 40 contracts,
and we decided to use those in our evaluation. 
Starting with those 40 contracts, we removed duplicates with trivial
syntactic differences,
resulting in a total of 25 unique contracts (Table~\ref{table:zeus}).
Thus, the objective of our evaluation is to run \verismart~and \smtchecker~on the 25 contracts to see
how many of them can be successfully analyzed by \verismart~and \smtchecker~without
false alarms. We ran \smtchecker~with the default setting. 


\myparagraph{Results} 
\newcommand{\ropsten}[2]{\href{https://ropsten.etherscan.io/address/#2\#code}{\footnotesize #1}}
\newcommand{\etherscan}[2]{\href{https://etherscan.io/address/#2\#code}{\footnotesize #1}}


\begin{table}[t]
\begin{footnotesize}
\caption{Evaluation on the \zeus~dataset.
 \Verified:  a tool detects all bugs without false positives
   (\cmark: success, \xmark: failure)
 }\label{table:zeus}
 \setlength{\tabcolsep}{0.23em}
\begin{tabular}{|lrr |rrc |rrc |c|} 
\hline
\multirow{2}{*}{\textsf{No.}} & \multicolumn{1}{c}{\multirow{2}{*}{\textsf{LOC}}} & \multirow{2}{*}{\Q} &
\multicolumn{3}{c|}{\verismart} & \multicolumn{3}{c|}{\smtchecker~\cite{DBLP:conf/isola/AltR18}} & \zeus~\cite{DBLP:conf/ndss/KalraGDS18}  \\ \cline{4-10}
&\multicolumn{2}{c|}{}&  \Alarms  & \FP & \Verified & \Alarms & \FP& \Verified & \Verified \\ \hline
\ropsten{\#1}{0xc98f5c1b3b783794e646a8a29e2916668b7d9606}  	     & 42     & 3  		& 0 & 0 &  \cmark&	 	3  & 3  & \xmark& \xmark\\
\etherscan{\#2}{0xa1b43b46befb2387d2df46cde82c3d454ef33c66} 	     & 78     & 2  		& 1 & 0 & \cmark&	 	2  & 1 & \xmark & \xmark \\
\etherscan{\#3}{0xd41f3b51e0c2d825a1178582d27c84dbfe48d1af} 	     & 75     & 7  		& 2 & 0 & \cmark& 		7  & 5  & \xmark& \xmark\\
\etherscan{\#4}{0x8a70cf25cf32e728be9e30c20b2781f60cb0ed6d} 	     & 70     & 7  		& 0 & 0 & \cmark& 		7  & 7  & \xmark& \xmark\\
\ropsten{\#5}{0xcfe185ce294b443c16dd89f00527d8b25c45bf9d}	     & 103   & 8  		& 0 & 0 & \cmark& 		6  & 6 & \xmark & \xmark \\
\etherscan{\#6}{0xd9f7cd813983bd89d18015cc3022f7b9b97d26d4} 	     & 141   & 5  		& 2 & 0 & \cmark& 		\multicolumn{3}{c|}{internal error} & \xmark\\
\ropsten{\#7}{0x218e5ea7e104385b0b91097519dfde91f15613c7} 	      & 74     & 6 		& 1 & 0 & \cmark& 		6  & 5 & \xmark& \xmark\\
\etherscan{\#8}{0x4993CB95c7443bdC06155c5f5688Be9D8f6999a5}     & 84     & 6 		& 0 & 0 & \cmark& 		4  & 4  & \xmark& \xmark  \\
\ropsten{\#9}{0xf48cf5ad04afa369fe1ae599a8f3699c712b0352} 	      & 82    & 6  		& 0 & 0 & \cmark& 		6  & 6  & \xmark& \xmark\\
\etherscan{\#10}{0x168296bb09e24a88805cb9c33356536b980d3fc5}     & 99    & 2  	& 1 & 0 & \cmark& 		\multicolumn{3}{c|}{internal error} & \xmark\\
\etherscan{\#11}{0x8c65898cceaa73209f579653fa5523b7b13972bd}      &171  & 15 		& 9 & 0 & \cmark&	 	\multicolumn{3}{c|}{internal error} & \xmark\\
\etherscan{\#12}{0xe57a41170f18fab3248d623f06bd92b32260fae2} 	      & 139  & 7  		& 0 & 0 & \cmark& 		\multicolumn{3}{c|}{internal error} & \xmark\\
\etherscan{\#13}{0xE01B770235Bc5db653604e5519F048dF54490B5f}  & 139  & 7  		& 0 & 0 & \cmark& 		\multicolumn{3}{c|}{internal error} & \xmark\\
\etherscan{\#14}{0xd23f2533B726C9Cb1Fb9ed109b82e5A8F01c881e}  & 139  & 7  		& 0 & 0 & \cmark& 		\multicolumn{3}{c|}{internal error} & \xmark\\
\etherscan{\#15}{0x3ff8c78e266395d08f41ef1631391f0050d48081}  	      &  139  & 7  	& 0 & 0 & \cmark& 		\multicolumn{3}{c|}{internal error} & \xmark\\
\etherscan{\#16}{0x45d147c800d401350b24fc1cd5fbc98040b177c8}      &  141  & 16	& 10& 0 & \cmark& 		\multicolumn{3}{c|}{internal error} & \xmark\\
\etherscan{\#17}{0x3B6d241e1b38776C2eFE944E7012239ed59334c1}  & 153  & 5 		& 0  & 0 & \cmark& 		\multicolumn{3}{c|}{internal error} & \xmark\\
\etherscan{\#18}{0x873c58020bcb114b4fea456cef93aaf58e8e305d}        &  139  & 7 	& 0 & 0 & \cmark& 		\multicolumn{3}{c|}{internal error} & \xmark\\
\etherscan{\#19}{0x08711d3b02c8758f2fb3ab4e80228418a7f8e39c}  	& 113  & 4  	& 0 & 0 & \cmark& 		4  & 4  & \xmark & \xmark \\
\etherscan{\#20}{0xd7bf41bbc8979b3821851b871f055f4ae62b2299}  	&  40	   & 3 	& 0 & 0 & \cmark& 		3  & 3  & \xmark& \xmark\\
\etherscan{\#21}{0x8a772004af0b8fca5e7093c6f277ba7b0e8fa97a}  	& 59    & 3  	& 0 & 0 & \cmark& 		\multicolumn{3}{c|}{internal error} & \xmark\\
\ropsten{\#22}{0x8d2c532d7d211816a2807a411f947b211569b68c}  		& 28    & 3  	& 1 &  0 & \cmark& 		1  & 0  & \cmark& \xmark\\
\ropsten{\#23}{0xeb41d678879c735f22fce499d891d44c288829ea}  		& 19    & 3  	& 0 & 0 & \cmark& 		3  & 3  & \xmark& \xmark\\
\etherscan{\#24}{0xcd3e727275bc2f511822dc9a26bd7b0bbf161784} 	&  457  & 30 	& 13& 6 & \xmark& 		\multicolumn{3}{c|}{internal error} & \xmark\\
\etherscan{\#25}{0xDea48D521832780f5e437F7f744c94d2CdA85Af9}  	& 17	   & 3  	& 0 & 0 & \cmark& 		3  & 3  & \xmark& \xmark\\ \hline
\multicolumn{1}{|c}{\multirow{2}{*}{\textbf{Total}}} & \multirow{2}{*}{2741} & \multirow{2}{*}{172}& \multirow{2}{*}{40}& \multirow{2}{*}{6} & \cmark:24 & \multirow{2}{*}{55} & \multirow{2}{*}{50}& \cmark:{ }{ }1  & \cmark:{ }0  \\
  			   & 				    &	   			   &    			 &				          & \xmark{ }: 1 &      		     &  			   &    			   	\xmark:{ }12 & \xmark:25 \\ \hline
\end{tabular}		
\end{footnotesize}
\vspace{-1em}
\end{table}

Table~\ref{table:zeus} shows
the evaluation results on the \zeus~dataset.  For each contract, the
table shows the number of alarms (\Alarms), the number of false
positives (\FP) produced by \verismart~and \smtchecker.
The column \textsf{Verified}
indicates whether each tool detected all bugs without false
positives (\cmark: success, \xmark: failure).

The results show that \verismart~successfully addresses limitations of \zeus~and \smtchecker. 
The 25 contracts contain 172 arithmetic operations, where
\verismart~pointed out 40 operations as potential bugs. We have
manually checked that 34 out of total alarms are true positives.
In benchmark \#24, \verismart~produced 6 false positives
due to unsupported invariants (quantified invariants and compound
invariants, Section~\ref{sec:generator}), and
imprecise function call analysis.
We manually checked that the remaining 132 (=172-40) queries proven to be safe by
\verismart~are actually true negatives.
By contrast, according to the publicly available data~\cite{zeus-report},
\zeus~produces at least one false positives for each
contract in Table~\ref{table:zeus} (i.e., $\ge 25$ false alarms in total). 
\smtchecker~could only analyze 13 contracts as it
raised internal errors for the other 12 contracts, which is due to its immature support of Solidity syntax~\cite{smtchecker-error}.  
Among 61 operations from 13 contracts, \smtchecker~succeeded to detect all 5 bugs in them thanks to its exhaustive verification approach.
However, it reported 55 alarms in total, of which 50 are false positives. 
In terms of efficiency, \smtchecker~took about 1 second per contract and 
\verismart~took about 20 seconds per contract. 



\myparagraph{Importance of Transaction Invariants}
The key enabler for high precision was the ability of \verismart~to leverage transaction invariants. 
We also ran \verismart~without inferring transaction invariants (i.e., using $\true$ as transaction invariants);
without transaction invariants, \verismart~fails to verify 17 out of 25 contracts.

\subsection{Case Study: Application to Other Types of Vulnerabilities}\label{sec:checker}

\verismart~can be used for analyzing other safety properties as well. 
To show this, we applied \verismart~to finding bugs related to access control, where security-sensitive variables can be manipulated by anyone for malicious use.  
For example, consider the code snippet adapted from the EtherCartel contract for
crypto idle game (CVE 2018-11329):
\begin{lstlisting}[numbers=none]
function DrugDealer() public { ceoAddr = msg.sender; }
function buyDrugs () public payable {
  ceoAddr.transfer(msg.value); // send Ether to ceoAddr
  drugs[msg.sender] += ...; // buy drugs by paying Ether
}
\end{lstlisting}
Observe that the address-typed variable \texttt{ceoAddr},
the beneficiary of Ether, 
can be taken by anyone who calls the function \texttt{DrugDealer}.
If an attacker becomes the beneficiary by calling \texttt{DrugDealer},
the attacker might illegally take some digital assets whenever benign users
buy some digital assets (i.e., drugs) by calling \texttt{buyDrugs}
where \texttt{transfer} in it is a built-in function that sends Ether to \texttt{ceoAddr}.
This vulnerability was exploited in about 1 hour after 
deployment~\cite{ceoAnyone}.

To detect this bug, we used \verismart~as follows. 
First, we specified safety properties by automatically generating the 
assertion \texttt{assert(msg.sender==addr)} right before each assignment of the form \texttt{addr=...;}, where \texttt{addr} is a global address-typed variable
which is often security-sensitive (excluding assignments in constructors, which typically set the contract owners).
Next, we ran \verismart~without any modification of its verification algorithm. 
With this simple extension, \verismart~worked effectively; it not only detected all
 known CVE vulnerabilities (2018-10666, 2018-10705, 2018-11329)
but also proved the absence of this bug scenario for
55 contracts out of 60 from Table~\ref{table:main}.
\verismart~could not prove safety of the remaining 5 contracts
due to the imprecise specification described above. 
 \subsection{Threats to Validity}\label{sec:threat}
We summarize limitations of our evaluation and consequent threats to validity. 
 Firstly, the benchmark contracts that we used
(60 CVE dataset + 25 \zeus~dataset) might not be representative
although we made effort to avoid bias in the datasets
(e.g., removal of duplicates). 
 Secondly, the performance of
\verismart~may vary depending on the performance of
the off-the-shelf SMT solver (i.e., Z3) used internally
or timeout options used in the experiments.
 Thirdly, we did not study the exploitability of bugs in this paper and did not
compare \verismart~ and other tools in this regard. Thus, the results
may be different if those tools are evaluated with 
exploitability in mind.
Lastly, although we did our best, we realized that manually classifying static analysis alarms into true or false
positives is extremely challenging and the classification can be
even subjective in a few cases. 







\section{Related Work}\label{sec:related}

In this section, we place our work in the literature and clarify our contributions regarding existing works. Section~\ref{sec:related1} compares our work with existing smart contract analyses. Section~\ref{sec:related2} discusses verification techniques for other domains. 

\subsection{Analyzing Smart Contracts}\label{sec:related1}

Compared to existing techniques for analyzing smart contracts~\cite{oyente,oyentetool, mythril,Nikolic:2018:FGP:3274694.3274743, DBLP:conf/acsac/TorresSS18, gasper, gastap,Grossman:2017:ODE:3177123.3158136,reguard,DBLP:conf/isola/AltR18,DBLP:conf/ndss/KalraGDS18,
DBLP:conf/ccs/TsankovDDGBV18, Grech:2018:MSO:3288538.3276486,
wh3-solidity, Hirai2017, Bhargavan2016, Grishchenko2018,Amani2018, DBLP:journals/corr/abs-1812-08829}, 
\verismart~is unique in that it achieves full automation, high precision, and high recall at the same time. 
%
%
Below, we classify existing approaches into
fully automated and semi-automated approaches. 


\myparagraph{Fully Automated Approaches}

\verismart~belongs to the class of fully automated tools based on static or dynamic program analysis techniques that require no manual effort and can be used by end-users
who lack expertise in formal verification.
Instead,  these approaches  focus
on relatively simple safety properties (e.g., overflows). 



One popular approach is bug-finders based on symbolic
execution or fuzz testing. For example,
\oyente~\cite{oyente,oyentetool}, \mythril~\cite{mythril},
\osiris~\cite{DBLP:conf/acsac/TorresSS18}, \manticore~\cite{manticore}
and
\maian~\cite{Nikolic:2018:FGP:3274694.3274743}~discover bugs by
symbolically executing EVM bytecode. \oyente~is the first such tool
for Ethereum smart contracts, which detects various bug patterns including arithmetic bugs. 
\mythril~is also a well-known
open-sourced tool for detecting a variety of bugs by performing
symbolic execution.
\osiris~\cite{DBLP:conf/acsac/TorresSS18} is a
tool that is specially designed for detecting arithmetic bugs. 
 \maian~\cite{Nikolic:2018:FGP:3274694.3274743} focuses on finding violations of trace properties.
\gasper~\cite{gasper} uses symbolic execution to identify gas-costly
programming patterns.  \reguard~\cite{reguard} and
ContractFuzzer~\cite{DBLP:conf/kbse/0001LC18} use fuzz testing to
detect common security vulnerabilities. 
Although symbolic execution and fuzz testing are
effective for finding bugs, they inevitably miss critical vulnerabilities, 
which is
particularly undesirable for safety-critical software like smart
contracts.


Other approaches are verifiers that perform exhaustive analyses based on static analysis or automatic program verification
techniques. 
\zeus~\cite{DBLP:conf/ndss/KalraGDS18} is a sound static analyzer that
can detect arithmetic bugs or prove their absence.  
\zeus~leverages abstract interpretation and software model checking~\cite{DBLP:conf/cav/GurfinkelKKN15}.
\smtchecker~\cite{DBLP:conf/isola/AltR18} is the ``official'' verifier for Solidity developed by the Ehtereum Foundation. 
Its primarily goal is to verify the absence of
arithmetic bugs such as integer over/underflows and
division-by-zeros~\cite{DBLP:conf/isola/AltR18} by performing SMT-based bounded verification. 
Unlike \verismart, \zeus~and \smtchecker~lack inter-transactional reasoning and this is currently considered a key limitation of these tools~\cite{DBLP:conf/ndss/KalraGDS18,DBLP:conf/isola/AltR18}. 

\securify~\cite{DBLP:conf/ccs/TsankovDDGBV18},
MadMax~\cite{Grech:2018:MSO:3288538.3276486}, and Vandal~\cite{vandal}
use declarative static analysis techniques based on
Datalog~\cite{Bravenboer2009}.  
Besides their inability to infer transaction invariants, 
one common drawback of Datalog-based analyzers is that they cannot
describe general classes of (in particular, numerical) static analyses
and is inappropriate for finding arithmetic bugs. 


\myparagraph{Semi-Automated Approaches}


Semi-automated tools for formally specifying and verifying smart
contracts have different goals. 
These approaches can 
prove a wide range of functional properties at the expense of full automation; they
require users to manually provide specifications or invariants. 

Hirai~\cite{Hirai2017} formalizes the Ethereum Virtual Machine (EVM)
and provides a way to prove safety properties of smart contracts in
interactive theorem provers such as Isabelle/HOL~\cite{Nipkow:2002:IPA:1791547}.  Bharagavan
et al.~\cite{Bhargavan2016} provide a framework for formally
specifying and verifying functional correctness of smart contracts
using the F* proof assistant~\cite{fstar}.  Grishchenko et al.~\cite{Grishchenko2018} also use
F* to formalize small-step semantics of EVM bytecode and express a
number of security properties of smart contracts.  Hildenbrandt et
al.~\cite{kevm} define formal semantics of EVM using the K
framework~\cite{rosu-serbanuta-2010-jlap}.  Amani et al.~\cite{Amani2018} formalize EVM in
Isabelle/HOL and provide a program logic for reasoning about smart
contracts.
Lahiri et al.~\cite{DBLP:journals/corr/abs-1812-08829} describe an
approach for formal specification and verification of smart
contracts, where the primary goal is to take a high-level specification expressed
by a state machine and
to verify that the implementation meets the specification.





\myparagraph{Manual Safety Checking}
Some techniques~(e.g., SafeMath~\cite{safemath}) depend on manual annotation of programs to prevent bugs, which has two drawbacks. 
First, manual annotation
is error-prone, hardly exhaustive, and sometimes
not recommended (e.g., decreasing readability, unnecessary waste of gas fees).
As a result, many smart contracts do not perform manual safety checking
exhaustively~\cite{DBLP:conf/acsac/TorresSS18, DBLP:conf/ndss/KalraGDS18}.  
Second, verification prevents bugs at compile time
so that they can be fixed before deployment, but manual checking detects bugs only at runtime.

\subsection{Analyzing Arithmetic Safety of Traditional Programs}\label{sec:related2}

Ensuring arithmetic safety has been studied extensively in the program analysis and verification communities~\cite{DBLP:conf/pldi/BlanchetCCFMMMR03, DBLP:journals/fmsd/CousotCFMMR09,sparrow,DBLP:journals/fac/KirchnerKPSY15, Frama-C,
DBLP:conf/osdi/WangCJZK12,DBLP:conf/ndss/WangWLZ09,DBLP:conf/uss/MolnarLW09,DBLP:conf/uss/MolnarLW09,moy2009modular,DBLP:conf/asplos/Sidiroglou-Douskos15}.
Our work differs from them in two ways. First, we focus on smart contracts and provide 
a domain-specific algorithm.
Second, to our knowledge, our CEGIS-style algorithm for verifying arithmetic safety is also new in this general context. 
%


Astr{\'{e}}e~\cite{DBLP:conf/pldi/BlanchetCCFMMMR03, DBLP:journals/fmsd/CousotCFMMR09} is a domain-specific static analyzer tailored to flight-control software. 
Sparrow~\cite{sparrow} and
Frama-C~\cite{DBLP:journals/fac/KirchnerKPSY15, Frama-C} are domain-unaware
static analyzers for C programs.
Astr{\'{e}}e, Sparrow, and Frama-C are based on abstract interpretation~\cite{Cousot:1977:AIU:512950.512973,Cousot:1979:SDP:567752.567778}. 
Instead, we use a CEGIS-style algorithm because existing abstract domains such as intervals~\cite{Cousot:1977:AIU:512950.512973} and octagons~\cite{DBLP:journals/lisp/Mine06} cannot capture domain-specific invariants (e.g., {\sf sum}) of smart contracts. Furthermore, abstract interpretation cannot infer invariants that are useful in practice but not inductive with respect to their abstract semantics. 
While our approach is similar to the existing CEGIS approaches 
(e.g.,~\cite{Solar-Lezama:2006:CSF:1168917.1168907,Udupa:2013:TSP:2491956.2462174,Solar-Lezama:2008:PSS:1714168}), to the best of our knowledge,
its application to arithmetic safety verification has not been studied.
Bounded verification approaches (e.g.,~\cite{DBLP:conf/tacas/ClarkeKL04,DBLP:conf/vstte/ClochardFP15}) are different from our work as we perform unbounded verification.
Our work is different from symbolic execution-based techniques~\cite{DBLP:conf/osdi/WangCJZK12,DBLP:conf/ndss/WangWLZ09,DBLP:conf/uss/MolnarLW09,DBLP:conf/uss/MolnarLW09,moy2009modular,DBLP:conf/asplos/Sidiroglou-Douskos15}
or unsound static analysis~\cite{Sarkar:2007:FSA:1332044.1332098,DBLP:conf/dimva/CeesayZGLB06}, as we~aim to detect all bugs. 
A few techniques aim to fix integer overflow bugs~\cite{DBLP:conf/popl/LongSKR14,Coker:2013:PTF:2486788.2486892,Cheng:2017:IAI:3155562.3155693}, which may introduce unwanted changes in programs though useful.

\section{Conclusion}

As smart contracts are safety-critical, formally verifying their
correctness is of the greatest importance. In this paper, we presented
a new and powerful verification algorithm for smart
contracts. Its central feature is the ability to automatically infer
hidden, in particular transaction, invariants of smart contracts and
leverage them during the verification process. 
We implemented the
algorithm in a tool, \verismart, for verifying arithmetic safety of
Ethereum smart contracts and
demonstrate its effectiveness on real-world smart contracts in
comparison with existing safety analyzers. 
Our work shows a common yet significant shortcoming of existing
approaches (i.e., inability to infer and use transaction invariants)
and sheds light on the future development of automated tools for analyzing
smart contracts.


\section*{Acknowledgment}
We thank Junhee Lee and Minseok Jeon for their valuable comments on 
Proposition~\ref{prop:invalidity-checking} and Appendix~\ref{sec:sum}.
This work was supported by Institute of Information \& communications
Technology Planning \& Evaluation(IITP) grant funded by the Korea
government(MSIT) (No.2019-0-01697, Development of Automated
Vulnerability Discovery Technologies for Blockchain Platform Security
and No.2019-0-00099, Formal Specification of Smart Contract).

\bibliographystyle{IEEEtran}
\bibliography{ref}

\appendix
\myappendix{Preprocessing of Verification Conditions}
\label{sec:sum}
Given a basic path $p$,
let $F$ be a verification condition
(either an inductiveness condition, i.e., $F = \genvc(p).1$,
or a safety condition, i.e., $F \in \genvc(p).2$)
that
contains equalities of the form $\mysigma(x)=e$ for some
mapping variable $x$ and expression $e$.
For simplicity, we assume that 
$F$ does not contain primed instances (e.g., $x'$, $x''$) of the mapping variable $x$. 
Let $I$ be the set of variables in $F$
used as indices of $x$.
Then, we replace each equality $\mysigma(x) = e$ by $G$ as follows.
If $I = \emptyset$, we define $G$ to be $G_1 \land G_2$
where $G_1=(R_x = e)$, $G_2=B_x$
($R_x$ and $B_x$ are fresh variables, see Section~\ref{sec:solver}).
If $I = \myset{i}$ (i.e., $I$ is a singletone set), we define
$G$ to be $G_1 \land G_2$ where
$G_1 = (x[i] + R_x = e)$ and $G_2 = (x[i] + R_x \ge R_x \land B_x)$. 
Otherwise (i.e., $I = \myset{i_1,\dots, i_n}, n \ge 2$),
we define $G$ to be $G_1 \land G_2$ where
\begin{equation*}
\begin{scriptsize}
\begin{multlined}
G_1 = \\
\bigwedge_{\substack{a \in [1,m], \\ P_a= \\ \myset{\myset{i_1, \dots}, \cdots, \myset{i_k, \dots}}}} \Big(
(\bigwedge_{I_u \in P_a} (\bigwedge_{i,j \in I_{u}} i=j) 
\land
\bigwedge_{\substack{I_u, I_v \in P_{a}, \\ I_{u}\not=I_{v}}} (\bigwedge_{\substack{i \in I_{u}, \\ j\in I_{v}}} i\not=j)) \to \\
x[i_1] + \cdots + x[i_k] + R_x = e \Big)
 \end{multlined}
 \end{scriptsize}
\end{equation*}
and
\begin{equation*}
\begin{scriptsize}
\begin{multlined}
G_2 = \\
\bigwedge_{\substack{a \in [1,m], \\ P_a= \\ \myset{\myset{i_1, \dots}, \cdots, \myset{i_k, \dots}}}} \Big(
(\bigwedge_{I_u \in P_a} (\bigwedge_{i,j \in I_{u}} i=j) 
\land
\bigwedge_{\substack{I_u, I_v \in P_{a}, \\ I_{u}\not=I_{v}}} (\bigwedge_{\substack{i \in I_{u}, \\ j\in I_{v}}} i\not=j)) \to \\
H_{x,i,k}
\land 
x[i_1] + \cdots + x[i_k] + R_x \ge R_x) \Big) 
\land B_x.
\end{multlined}
\end{scriptsize}
\end{equation*}
$H_{x,i,k}$ is defined as $\true$ if $k=1$, and
defined as $\bigwedge_{c=2}^{k} x[i_1] + \cdots + x[i_c] \ge x[i_c]$ otherwise (i.e., $k\ge2$).
$P_1,\dots,P_m$ are all possible partitions of the index variable set $I$,
where a partition is a set of disjoint non-empty subsets of $I$ such
that the union of the subsets equals $I$.
For example, given $I=\myset{i,j}$, we have two partitions: $\myset{\myset{i,j}}$ and
$\myset{\myset{i}, \myset{j}}$. Also, given $I=\myset{i,j,k}$, we have five partitions:
$\myset{\myset{i,j,k}}$, $\myset{\myset{i}, \myset{j,k}}$, $\myset{\myset{j}, \myset{i,k}}$,
$\myset{\myset{k}, \myset{i, j}}$, and $\myset{\myset{i}, \myset{j}, \myset{k}}$.

Intuitively, $G_1$ asserts
that the sum of distinct elements of $x$ equals $e$,
and $G_2$ asserts
that overflows do not occur during computing the sum of the distinct elements.
More specifically, using the partitions of $I$, we first consider all possible
cases of (in)equalities among the variables in $I$;
for each partition $P_a =\myset{I_1,\dots,I_k}$ (where $\uplus_{1 \le i \le k} I_i = I$),
the variables in the same subsets have the same values
(i.e., $\bigwedge_{I_u \in P_a} (\bigwedge_{i,j \in I_u}  i=j)$),
and the variables in different subsets have different values
(i.e., $\bigwedge_{I_u, I_v \in P_a, I_u \not= I_v} (\bigwedge_{i \in I_u, j \in I_v} i\not=j)$).
Then, for each partition case, we generate constraints on the distinct elements of $x$ by
selecting an index variable from each subset.




\begin{example}
Given a basic path $p$,
suppose $F \in \genvc(p).2$ is given as follows:
\[
\mysigma (y) = 100 \land y[i] \ge v \to y[j] + v \ge y[j]
\]
In this case, the index variable set for $y$ is $I=\myset{i,j}$,
because $i$ and $j$ are used as indices in $y[i]$ and $y[j]$, respectively.
For $I$, we have two partitions $P_1=\myset{\myset{i,j}}$ and $P_2=\myset{\myset{i},\myset{j}}$,
and thus we consider two cases: $i=j$ from $P_1$ and $i\not=j$ from $P_2$.
Then, we replace $\mysigma(y)=100$ by $G_1 \land G_2$ where
$G_1$ is
\[
\begin{small}
\begin{array}{l}
 (i\not=j \to y[i] + y[j] + R_y = 100) \land 
 (i=j \to y[i] + R_y = 100)
 \end{array}
\end{small}
\]
and $G_2$ is 
\[
\begin{small}
\begin{array}{l}
(i\not=j \to y[i] + y[j] \ge y[i] \land y[i]+ y[j] + R_y \ge R_y) \land \\
 (i=j \to y[i] + R_y \ge R_y) \land B_y.
\end{array}
\end{small}
\]
Finally, by replacing $\mysigma(y)=100$ in $F$ by $G_1\land G_2$,
we obtain the following $F'$
\[
\begin{array}{l}
\big(
 (i\not=j \to y[i] + y[j] + R_y = 100) \land \\
 (i=j \to y[i] + R_y = 100) \land \\
(i\not=j \to y[i] + y[j] \ge y[i] \land y[i]+ y[j] + R_y \ge R_y) \land \\
 (i=j \to y[i] + R_y \ge R_y) \land B_y \land \\
 y[i] \ge v \big) \\
 \to y[j] + v \ge y[j]
\end{array}
\]
which is satisfiable iff the original formula $F$ is satisfiable. 
%
\end{example}

\myappendix{Proof of Proposition \ref{prop:invalidity-checking}}
\label{sec:invalidity-checking}

Proof by contradiction.
Assume $p \implies q$:
\begin{equation}\label{eq1}
\forall I. I \models \neg p \lor q.
\end{equation}
From condition (\ref{sitm2}) and (\ref{eq1}), we have
\begin{equation}\label{eq2}
I_p \models q
\end{equation}
where $I_p$ is an interpretation that makes the evaluation of $p$ $\true$ (i.e., $I_p \models p$).
From condition (\ref{sitm2}), condition (\ref{sitm3}), and (\ref{eq2}), we have a $x$-variant of $I_p$, denoted as $I'_p$, such that
\begin{equation}\label{eq3}
I'_p: I_p \vartriangleleft \myset{x \mapsto v} \models \neg q
\end{equation}
where $x \in \free(q) \setminus \free(p)$ and
$v\in D_{I_p} \setminus \myset{\alpha_{I_p}[x]}$.
Since $I_p \models p$ and $x \not\in \free(p)$,
\begin{equation}\label{eq4}
I'_p \models p.
\end{equation}
Combining~(\ref{eq3}) and~(\ref{eq4}), we have $I'_p \models \neg(\neg p \lor q)$,
which contradicts with the assumption~(\ref{eq1}).
Thus $p \centernot\implies q$.

\myappendix{More Examples of Validity Templates}
\label{sec:template-example}


We provide three more examples that are important for performance. 
We assume that formula $F$ below is in CNF (conjunctive normal form). 
We write $c \in F$ for indicating that $F$ has clause $c$. 
\begin{example}\label{ex:domain-specific-template1}
Consider a template
\[
\infer[n+ n \ge n]
{F \to x[q] + v \ge x[q]}
{\mysigma(x) = n \in F, ~ x[p] \ge v \in F}
\]
where $x$ is a mapping variable that maps address-typed index variables to
256-bit unsigned integer-typed variables,
$n$ is an integer constant (where $n+n$ does not overflow in unsigned 256-bit),
and $p$ and $q$ are address-typed variables. 
The template above states that, when $\mysigma(x) = n$ and $x[p] \ge v$ hold in the precondition $F$,
$x[q] + v \ge x[q]$ also holds for any index address-typed variable $q$.
For example, we can use the rule to check that the VC 
\[
\dots \land \mysigma(a) = 100 \land \dots \land a[i] \ge k \land \dots \to a[j] + k \ge a[j]
\]
is valid without preprocessing the formula and invoking an SMT solver. 
\end{example}


\begin{example} 
Consider a template:
\[
\infer[n+ n \ge n]
{F \to x[q] + v \ge x[q]}
{\mysigma(x) = y \in F, ~y=n \in F, ~x[p] \ge v \in F}
\]
where $x$ is a mapping variable that maps address-typed index variables to 256-bit unsigned integer-typed variables,
$y$ and $v$ are 256-bit unsigned integer-typed variables,
$n$ is an integer constant (where $n+n$ does not overflow in unsigned 256-bit),
and $p$ and $q$ are address-typed variables.
Note that the template above is similar to the one
in Example~\ref{ex:domain-specific-template1}, where
$\mysigma(x) = n$ is changed into a combination of $\mysigma(x) = y$ and $y=n$.
Using the template, we can prove the validity of the VC:
\[
\begin{array}{l}
\dots \land \mysigma(a) = b \land \dots \land b=100 \land \dots \land a[i] \ge k \land \dots  \\
\to a[j] + k \ge a[j]
\end{array}
\]
\end{example}


\begin{example} 
Consider a template:
\[
\infer [n_1 + n_2 \ge n_1]
{F \to n_1 + (x \% n_2) \ge n_1} 
{  }
\]
where $x$ is a 256-bit unsigned integer-typed variables, and
$n_1$ and $n_2$ are integer constants (where $n_1+n_2$ does not overflow in unsigned 256-bit).
Using the validity template above, we can
prove that $\dots \to 48 + (y \% 10) \ge 48$ is valid.
\end{example}




\end{document}